\documentclass[twocolumn,showpacs,preprintnumbers,amsmath,amssymb,pre]{revtex4}

\usepackage{graphicx}
\usepackage{dcolumn}
\usepackage{bm}
\usepackage{enumerate}
\usepackage{amsmath}
\usepackage{color}

\begin{document}

\title{Subdiffusion equation with fractional Caputo time derivative with respect to another function in modeling transition from ordinary subdiffusion to superdiffusion}

\author{Tadeusz Koszto{\l}owicz}
 \email{tadeusz.kosztolowicz@ujk.edu.pl}
 \affiliation{Institute of Physics, Jan Kochanowski University,\\
         Uniwersytecka 7, 25-406 Kielce, Poland}

\date{\today}

\begin{abstract}
We use a subdiffusion equation with fractional Caputo time derivative with respect to another function $g$ ($g$--subdiffusion equation) to describe a smooth transition from ordinary subdiffusion to superdiffusion. Ordinary subdiffusion is described by the equation with the ``ordinary'' fractional Caputo time derivative, superdiffusion is described by the equation with a fractional Riesz type spatial derivative. We find the function $g$ for which the solution (Green's function, GF) to the $g$--subdiffusion equation takes the form of GF for ordinary subdiffusion in the limit of small time and GF for superdiffusion in the limit of long time. To solve the $g$--subdiffusion equation we use the $g$--Laplace transform method. It is shown that the scaling properties of the GF for $g$--subdiffusion and the GF for superdiffusion are the same in the long time limit. We conclude that for a sufficiently long time the $g$--subdiffusion equation describes superdiffusion well, despite a different stochastic interpretation of the processes. Then, paradoxically, a subdiffusion equation with a fractional time derivative describes superdiffusion. The superdiffusive effect is achieved here not by making anomalously long jumps by a diffusing particle, but by greatly increasing the particle jump frequency which is derived by means of the $g$--continuous time random walk model. The $g$--subdiffusion equation is shown to be quite general, it can be used in modeling of processes in which a kind of diffusion change continuously over time. In addition, some methods used in modeling of ordinary subdiffusion processes, such as the derivation of local boundary conditions at a thin partially permeable membrane, can be used to model $g$--subdiffusion processes, even if this process is interpreted as superdiffusion.
\end{abstract}

\maketitle

\section{Introduction\label{SecI}}

In the last few decades, various types of diffusion have been studied experimentally and theoretically. Fractional differential calculus has been widely used in the theoretical description of anomalous diffusion. The list of citations related to these issues is very long, as we mention here \cite{bg,ks,hughes,klages2008,wyss1986,schneider1989,mks,skb,sk,barkai2000,compte,mk,mk1,hilferanton,barkai2012}.
Most often, diffusion processes are characterized by a time evolution of the mean square displacement (MSD) of a diffusing particle $\sigma^2$. If $\sigma^2(t)\sim t^\beta$, $\beta\in(0,1)$ corresponds to ordinary subdiffusion, $\beta=1$ to normal diffusion, and $\beta>1$ to superdiffusion. For ultraslow diffusion (slow subdiffusion) $\sigma^2$ is controlled by a slowly varying function which, in practice, is a combination of logarithmic functions \cite{chechkin,denisov}. A more detailed description of diffusion models generating different functions $\sigma^2$ is in Refs. \cite{mjcb,csm1}.

Subdiffusion occurs in media in which the movement of particles is very hindered due to a complex structure of a medium. The examples are transport of some molecules in viscoelastic chromatin network \cite{lee2021}, porous media \cite{bijeljic2011}, living cells \cite{barkai2012}, transport of sugars in agarose gel \cite{kdm}, transport of water in aqueous sucrose glasses \cite{zorbist2011}, and antibiotics in bacterial biofilm \cite{km,kmwa}. Subdiffusion can also occur in a medium with normal diffusion near the membrane, which retains diffusing molecules for a very long time \cite{kd2021}.
Within the continuous time random walk (CTRW) model, the distribution $\psi$ of waiting time for a particle to jump $\Delta t$ has a heavy tail, $\psi(\Delta t)\sim 1/(\Delta t)^{\alpha+1}$ when $\Delta t\rightarrow\infty$, $\alpha\in(0,1)$; the average value of this time is infinite. The ordinary subdiffusion equation with the Riemann--Liouville time derivative of the order $1-\alpha$ or Caputo fractional time derivative of the order $\alpha$ is frequently used to describe subdiffusion, then $\sigma^2\sim t^\alpha$. For superdiffusion, long particle jumps can be performed with relatively high probability, a jump length distribution $\lambda$ has a heavy tail, $\lambda(\Delta x)\sim 1/|\Delta x|^{1+\gamma}$ when $|\Delta x|\rightarrow\infty$; in the following we consider the case of $\gamma\in(1,2)$. The second moment of the jump length is infinite, the process is described by a differential equation with a fractional Riesz type derivative of the order $\gamma$ with respect to a spatial variable. In this case the CTRW model provides $\sigma^2(t)=\kappa t^{2/\gamma}$, however, $\kappa$ is infinite (see Eqs. (\ref{eqV8}) and (\ref{eqV9}) presented later). Therefore, superdiffusion is often defined by the relation $\sigma^2(t)\sim t^{2/\gamma}$ only and the prefactor is usually not considered. Examples of superdiffusion are movement of endogeneous intracellular particles in some pathogens \cite{reverey2015}, of soil amebas on plastic or glass surfaces in liquid media \cite{levandowsky1997}, mussels movement \cite{dejager2011}, cell migration in some biological processes \cite{dieterich2008}, and diffusion in random velocity fields \cite{redner1989,zumofen1990,bouchaud1990,compte1998}. 

Differential equations mentioned above with constant parameters are used to describe diffusion in a homogeneous medium which properties does not change with time. However, diffusion processes may occur in systems in which diffusion parameters and even a type of diffusion evolves over time. The type of diffusion depends on the interaction of diffusing molecules with the environment and on a structure of the medium, both can change over time. Anomalous diffusion with evolving anomalous diffusion exponent has been observed in transport of colloidal particles between two parallel plates \cite{chakrabarty2015}, endogenous lipid granules in living yeast cells \cite{jeon2011}, microspheres in a living eukaryotic cell \cite{caspi2002} and in bacterial motion on small beads in a freely suspended soap film \cite{wu2000}. A change in the diffusion type can occur in the diffusion of passive molecules in the active bath where moving particles can affect the movement of passive molecules. Active swimmers can enhance diffusion of passive particles \cite{mino2011}. The change of diffusion of self-propelled particles and passive particles in an environment with motile microorgamisms is described in Ref. \cite{bechinger2016}. Active molecules can take energy from the environment and use it to make long jumps. Some diffusing molecules can use chemical reactions to achieve autonomous propulsion \cite{howse2007}. This mechanism leads to the process in which $\sigma^2$ evolves much faster than the linear function of time. The nature of diffusion may also change when the directed movement of a molecule over short time intervals is disturbed by a random change in the direction of the molecule's movement due to rotational diffusion.

In intracellular transport in most eukaryotic cells molecules diffuse through the filament network. When particle transport is carried out along filaments, the particles can move ballistically (i.e. with $\beta=2$). However, changing the orientation and polarization of the filaments may change the nature of diffusion \cite{ando2015}. Some microorganisms, such as bacteria, move more quickly in more viscous media. This is because the addition of a viscosity enhancer creates a quasi-rigid network to facilitate the transport of molecules \cite{berg1979,magariyama2002}. Thus, an increase in viscosity may paradoxically facilitate diffusion. Various bacterial defense mechanisms against the action of an antibiotic may hinder but also facilitate the diffusion of antibiotic molecules in the biofilm \cite{aot,mot}. Bacterial defense abilities evolve over time, which can change diffusion parameters \cite{km,kmwa}.

When the diffusion parameters are not constant, various equations have been used to describe the diffusion processes. The examples are subdiffusion equations with a fractional time derivative of the order depending on time and/or on a spatial variable \cite{chen2013,sun2019,yzw,lwc,roth,awad,awad2020,patnaik,sun2010,sun2009,sun2012,fedotov}, the superstatistics approach \cite{chechkin2002} in which certain distribution of diffusion parameters is assumed, subdiffusion equation with distributed fractional order derivative \cite{sandev2015}, and with a linear combination of fractional time derivatives of different orders where the time evolution of MSD is a linear combination of power functions with different exponents \cite{baz}. Diffusion of passive molecules in suspension of eukaryotic swimming microorganisms is describes by a probability density function which is a linear combination of Gaussian and exponential Laplace distributions; we mention that validity of this function has been checked by means of the scaling method \cite{leptos2009}. Distributed order of fractional derivative in subdiffusion equation can lead to delayed or accelerated subdiffusion \cite{chechkin2002,chechkin2008,orzel,eab,eab1}.

To extend the possibilities of modeling anomalous diffusion processes diffusion equations with various fractional derivatives have been used. We mention here Cattaneo--Hristov diffusion equation with Caputo--Fabrizio fractional derivarive \cite{hristov2017,caputo2015,santos2018}, Erdelyi--Kober fractional diffusion equation \cite{pagnini2012,pagnini2013}, equations with Antagana--Baleanu--Caputo and Antagana--Baleanu--Riemann--Liouville fractional derivatives \cite{atangana2016,sene2019}, $\psi$--Hilfer derivatives \cite{vieira}, equations involving fractional derivatives with kernels depending on the Mittag--Leffler function, the examples are a Wiman type \cite{liang2018} and a Prabhakar type fractional diffusion equations \cite{singh2019,stanislavsky2018,giusti2020}, see also \cite{luchko2020,baleanu2019,baleanu2020,yu2018,failla2020}. Another generalization of the anomalous diffusion equation is involving of fractional derivative with respect to another function $g$ ($g$--fractional derivative) in the equation \cite{abdelhedi2021,almeida2017,kilbas2006,sousa2018}. Examples are anomalous diffusion equations with the $g$--Caputo fractional derivative with respect to time \cite{samet2019,garra2018} and to a spatial variable \cite{bohaienko2020}. The subdiffusion equation with fractional $g$--Caputo time derivative has recently been used to describe the transition from ordinary subdiffusion to slow subdiffusion \cite{kd2021a}, the transition between the ordinary subdiffusion processes with different subdiffusion parameters (exponents) \cite{kd2022}, and subdiffusion of particles vanishing over time \cite{koszt2022}. The $g$--subdiffusion equation has been also used to describe diffusion of colistin molecules in a system consisting of densely packed gel beads immersed in water \cite{kdlwa2022}. It has been shown that this process cannot be described by the ordinary subdiffusion equation nor normal diffusion one, but the application of the $g$--subdiffusion equation gives good agreement of the experimental and theoretical results. A more difficult task is to develop a model that describes the transition between subdiffusion and superdiffusion, as these processes have a different interpretations and are described by equations with fractional time derivative and fractional spatial derivative, respectively.

Changing of a time scale in a diffusion model can lead to changes in diffusion parameters and/or in the type of diffusion \cite{hilferanton,kd2021a}. A timescale changing can be made by means of subordinated method \cite{ks,sokolov,feller,csm,dybiec,chechkin2021}. It is also generated by diffusing diffusivities where the diffusion coefficient evolves over time \cite{csm}, passages through the layered media \cite{carr}, and anomalous diffusion in an expanding medium \cite{levot}. The timescale changing can provide retarding and accelerating anomalous diffusions \cite{stanislavsky2020,stanislavsky2019}. In the $g$--subdiffusion process, the $g$ function changes the time scale with respect to ordinary subdiffusion (assuming that the subdiffusion parameters of both processes are the same). Contrary to the subordinated method, for $g$--subdiffusion a time variable is rescaled by a deterministic function $g$.

The normal diffusion, ordinary subdiffusion, and fractional superdiffusion equations have been derived from the standard CTRW model \cite{mk,ks,mks,barkai2000,compte,montroll1965}. To derive more general anomalous diffusion equations from the CTRW model (if possible), further modifications of CTRW model should be made.
The CTRW model with time distribution $\psi$ controlled by a three-parameters Mittag--Leffler function provides the subdiffusion equation with the fractional Prabhakar derivative \cite{sandev2018}; the equation describes transition subdiffusion process between different subdiffusion exponents. The $g$--subdiffusion equation can be derived from the modified CTRW model \cite{kd2021b} in which the special definition of convolution of functions has been applied. 

A frequently used method of analytical solving of diffusion equations with fractional derivatives, apart from the method of variable separation, is the integral transform method. For the ordinary subdiffusion equation with Caputo or Riemann--Liouville time derivatives, the ordinary Laplace transform method is effective. For diffusion equations with other fractional derivatives, other methods are more effective. For example, fractional Hilfer--Prabhakar and Cattaneo--Hristov diffusion equations can be solved by means of the Elzaki transform method \cite{singh2019}. An effective method for solving the $g$--subdiffusion equation is the $g$--Laplace transform method \cite{kd2021a,kd2021b}.

Our considerations focus on the transition from subdiffusion to superdiffusion. Such a transition has been observed in dust diffusion in a flowing plasma in the presence of a moderate magnetic field \cite{dutta2021}. Diffusion of molecules, whose movement is limited, when the strength of the noise changes shows the subdiffusion--superdiffusion transition \cite{ribeiro2016}. The transition can be facilitated in a system where both processes coexist. Small changes in the parameters affecting diffusion, such as temperature or viscosity, can cause the transition. Simulations show the possibility of coexistence of subdiffusion and superdiffusion in heterogeneous media, especially near the points where diffusion parameters are discontinuous \cite{zhang2009}. Diffusion of particles interacting via Yukava potential in a two dimensional system also shows both subdiffusive and superdiffusive behaviour with time varying anomalous diffusion exponent \cite{ott2009}. We mention that the scaling method is helpful for identifying superdiffusion, see \cite{osp1985,zumofen1990,bouchaud1990,cecconi2022,kelly2019}.

We propose a model of a smooth transition from subdiffusion described by the equation with the ``ordinary'' fractional Caputo time derivative (hereinafter referred to as ordinary subdiffusion) to superdiffusion described by the equation with the fractional Riesz type spatial derivative (refereed to as fractional superdiffusion). The model is based on the $g$--subdiffusion equation with the fractional time Caputo derivative with respect to another function $g$. We consider the process in a one-dimensional homogeneous and isotropic system. 

The paper is organized as follows. In Sec. \ref{SecII} there are presented definitions and methods of computing functions describing diffusion, used in further considerations, in particular the Green function and the function $F$ which is interpreted as the first passage time distribution for subdiffusion and quasi-first passage time distribution for superdiffusion. The ordinary subdiffusion equation, $g$--subdiffusion equation, and fractional superdiffusion equation are described in Secs. \ref{SecIII}, \ref{SecIV}, and \ref{SecV}, respectively. In Sec. \ref{SecVI} we present a model of a smooth transition from ordinary subdiffusion, described by the equation with a fractional time derivative, to superdiffusion described by the equation with a fractional spatial derivative. The process is described by the fractional $g$--subdiffusion equation with an appropriately chosen function $g$. A smooth transition between the processes means that the Green's function describing this transition is smooth, i.e. its derivative exists and is continuous in the entire domain. The scaling properties of Green's function describing the transition process are considered in Sec. \ref{SecVII}. We assume that if the scaling properties of this function are consistent with the scaling properties of the Green's function for fractional superdiffusion, $g$--subdiffusion can be treated as superdiffusion. The stochastic interpretation of the transition process within the modified CTRW model is discussed in Sec. \ref{SecVIII}. Using this model, the time evolutions of the average jumps number and the jumps frequency of a diffusing particle are derived. These functions are used to interpret the considered $g$--subdiffusion process. Final remarks are presented in Sec. \ref{SecIX}. Some details of the calculations and properties of the H--Fox function are shown in Appendix.

\section{Functions that describe diffusion\label{SecII}}

We define the Green's function $P$ and the function $F$ which is the probability distribution of the time that a particle pass a selected point for first time for ordinary subdiffusion and $g$--subdiffusion. Despite interpretation difficulties, which will be explained later, in further considerations this function is also used for fractional superdiffusion.

The Green's function is defined here as a solution to a diffusion equation for the boundary conditions 
\begin{equation}\label{eqII1}
P(\pm\infty,t|x_0)=0,
\end{equation}
and the initial condition
\begin{equation}\label{eqII2}
P(x,0|x_0)=\delta(x-x_0),
\end{equation}
where $\delta$ denotes the delta--Dirac function.
The Green's function is interpreted as a probability density of finding a diffusing particle at point $x$ at time $t$, $x_0$ is the initial particle position. If the molecules diffuse independently of each other, their concentration $C(x,t)$ at a point $x$ at time $t$ can be calculated using the integral formula
\begin{equation}\label{eqII3}
C(x,t)=\int_{-\infty}^\infty C(x_0,0)P(x,t|x_0)dx_0,
\end{equation}
$C(x_0,0)$ is the initial substance concentration.
In unbounded homogeneous system without a bias, the Green's function is symmetrical with respect to $x-x_0$ and is invariant with translation. Then, the Green's function depends on the distance between the points $x_0$ and $x$ only and can be written as
\begin{equation}\label{eqII4} 
P(x,t|x_0)\equiv P(|x-x_0|,t|0).
\end{equation} 
Due to Eq. (\ref{eqII4}), the mean value of a particle position is $\left\langle x\right\rangle=\int_{-\infty}^\infty xP(x,t|x_0)dx=x_0$ and the mean square displacement (MSD) is
\begin{equation}\label{eqII5}
\sigma^2(t)=2\int_0^\infty x^2 P(x,t|0)dx.
\end{equation}

Determination of the first passage time probability density for ordinary subdiffusion and normal diffusion is as follows. Let $F(t;x_0,x_M)$ be a probability density that the particle located at $x_0$ at the initial moment $t=0$ will pass the point $x_M$ first time at time $t$; we assume $x_0<x_M$.
The probability that the particle leaves the region $I=(-\infty,x_M)$ first time in the time interval $(t,t+\Delta t)$, where $\Delta t$ is assumed to be small, is \cite{redner}
\begin{equation}\label{eqII6}
F(t;x_0,x_M)\Delta t=R(t;x_0,x_M)-R(t+\Delta t;x_0,x_M),
\end{equation}
where $R(t;x_0,x_M)$ denotes the probability that the particle would not have passed the point $x_M$ by the time $t$. The function $R$ can be calculated by means of the formula
\begin{equation}\label{eqII7}
R(t;x_0,x_M)=\int_{-\infty}^{x_M} P_{\rm abs}(x,t|x_0)dx,
\end{equation}
where $P_{\rm abs}(x,t|x_0)$ is a probability of finding the particle in the region $I$ in a system in which a fully absorbing wall is located at $x_M$ \cite{redner}. The commonly used boundary condition at the absorbing wall is
\begin{equation}\label{eqII8}
P_{\rm abs}(x_M,t|x_0)=0. 
\end{equation}
The Green's function for a system with a fully absorbing wall can be found by means of the method of images \cite{feller,chandra}, which for $x,x_0<x_M$  gives
\begin{equation}\label{eqII9}
P_{\rm abs}(x,t|x_0)=P(x,t|x_0)-P(x,t|2x_M-x_0).
\end{equation}
We mention that the method of images has been used to find the Green's function in a system with fully absorbing wall for ordinary subdiffusion \cite{metzler2000} and for $g$--subdiffusion \cite{koszt2022}.
From Eqs. (\ref{eqII4}), (\ref{eqII7}), and (\ref{eqII9}) we get
\begin{equation}\label{eqII10}
R(t;x_0,x_M)=2\int_0^{x_M-x_0} P(x,t|0)dx.
\end{equation}
Taking the limit of $\Delta t\rightarrow 0$ we obtain for $t>0$
\begin{equation}\label{eqII11}
F(t;x_0,x_M)=-\frac{dR(t;x_0,x_M)}{dt}.
\end{equation}
From Eqs. (\ref{eqII10}) and (\ref{eqII11}) we get
\begin{equation}\label{eqII12}
F(t;x_0,x_M)=-2\frac{d}{dt}\int_0^{x_M-x_0} P(x,t|0)dx.
\end{equation}
For $t< 0$ we put $F(t;x_0,x_M)\equiv 0$. 

The function $F$ is the first passage time probability density for ordinary subdiffusion and normal diffusion. However, for fractional superdiffusion this function does not satisfy the Sparre-Andersen theorem that for a symmetric discrete--time random walk there is $F\sim 1/n^{3/2}$ when $n\rightarrow\infty$, where $n$ is a number of particle jumps \cite{ks,chechkin2003,palyulin2019}. If the average waiting time for a particle to jump is finite, which is the case for fractional superdiffusion and normal diffusion, then $t\sim n$ and $F\sim 1/t^{3/2}$ when $t\rightarrow\infty$. However, for fractional superdiffusion Eq. (\ref{eqII12}) provides $F\sim 1/t^{1+1/\gamma}$, $\gamma\in(1,2)$, in the long time limit, see Ref. \cite{chechkin2003,palyulin2019,dybiec2009} and Eq. (\ref{eqV6}) presented later in this paper. This result is interpreted that the method of images is not applicable to fractional superdiffusion. Since the method provides Eq. (\ref{eqII8}), the boundary condition at the absorbing wall Eq. (\ref{eqII8}) is not valid for fractional superdiffusion \cite{chechkin2003}. Despite the difficulties of interpretation, the function $F$ defined by Eq. (\ref{eqII12}) shows interesting asymptotic properties of the fractional superdiffusion model and will be used in our considerations. We call it the distribution of quasi-first passage time for superdiffusion and denote it with the symbol $F_\gamma$.

\section{Ordinary subdiffusion equation\label{SecIII}}

Functions describing ordinary subdiffusion are denoted by the index $\alpha$. The ordinary subdiffusion equation of the order $\alpha\in(0,1)$ with Caputo fractional time derivative is 
\begin{equation}\label{eqIII1}
\frac{^C \partial^\alpha P_\alpha(x,t|x_0)}{\partial t^\alpha}=D_\alpha\frac{\partial^2 P_\alpha(x,t|x_0)}{\partial x^2},
\end{equation}
where the ordinary Caputo fractional derivative is defined for $\alpha\in(0,1)$ as 
\begin{equation}\label{eqIII2}
\frac{^Cd^\alpha f(t)}{dt^\alpha}=\frac{1}{\Gamma(1-\alpha)}\int_0^t (t-u)^{-\alpha}f'(u)du,
\end{equation}
where $f'(u)=df/du$, $\Gamma$ is the Gamma function, and the subdiffusion coefficient $D_\alpha$ is given in the units of ${\rm m^2/s^\alpha}$. The solution to Eq. (\ref{eqIII1}) can be find by means of the ordinary Laplace transform method. The ordinary Laplace transform
\begin{equation}\label{eqIII3}
\mathcal{L}[f(t)](s)=\int_0^\infty {\rm e}^{-st} f(t)dt
\end{equation} 
has the property
\begin{equation}\label{eqIII4}
\mathcal{L}\left[\frac{^C d^\alpha f(t)}{dt^\alpha}\right](s)=s^\alpha\mathcal{L}[f(t)](s)-s^{\alpha-1}f(0),
\end{equation}
when $\alpha\in(0,1)$. Due to Eq. (\ref{eqIII4}), the ordinary subdiffusion equation in terms of the ordinary Laplace transform reads
\begin{eqnarray}\label{eqIII5}
s^\alpha \mathcal{L}[P_\alpha(x,t|x_0)](s)-s^{\alpha-1}P_\alpha(x,0|x_0)\\
=D_\alpha\frac{\partial^2 \mathcal{L}[P_\alpha(x,t|x_0)](s)}{\partial x^2}.\nonumber
\end{eqnarray}
The solution to Eq. (\ref{eqIII5}) for the boundary conditions $\mathcal{L}[P_\alpha(\pm\infty,t|x_0)](s)=0$ (see Eq. (\ref{eqII1})) and the initial condition Eq. (\ref{eqII2}) is
\begin{equation}\label{eqIII6}
\mathcal{L}[P_\alpha(x,t|x_0)](s)=\frac{1}{2\sqrt{D_\alpha}s^{1-\alpha/2}}\;{\rm e}^{-|x-x_0|\sqrt{\frac{s^\alpha}{D_\alpha}}}.
\end{equation}
Using the following relation \cite{tkoszt2004}
\begin{eqnarray}\label{eqIII7}
\mathcal{L}^{-1}\left[s^\nu {\rm e}^{-as^\beta}\right](t)\equiv \tilde{f}_{\nu,\beta}(t;a)\\
=\frac{1}{t^{1+\nu}}\sum_{j=0}^\infty \frac{1}{j!\Gamma(-\nu-\beta j)}\left(-\frac{a}{t^\beta}\right)^j,\nonumber
\end{eqnarray}
where $a,\beta>0$, $\tilde{f}_{\nu,\beta}$ is a special case of the H--Fox function (the Wright function), from Eq. (\ref{eqIII6}) we get (see \cite{mainardi1996})
\begin{eqnarray}\label{eqIII8}
P_\alpha(x,t|x_0)=\frac{1}{2\sqrt{D_\alpha}}\tilde{f}_{-1+\alpha/2,\alpha/2}\left(t;\frac{|x-x_0|}{\sqrt{D_\alpha}}\right)\\
=\frac{1}{2\sqrt{D_\alpha t^\alpha}}\sum_{j=0}^\infty \frac{1}{j!\Gamma(1-\alpha(j+1)/2)}\left(-\frac{|x-x_0|}{\sqrt{D_\alpha t^\alpha}}\right)^j.\nonumber
\end{eqnarray}
The above function is also called the Mainardi function \cite{pagnini2013}. We mention that the relation (\ref{eqIII7}) has been considered for negative values of $\nu$ in a series of papers by Mainardi et al., see for example \cite{mainardi1996,apelblat,mainardi2022}.

Eq. (\ref{eqIII6}) and the ordinary Laplace transform of Eq. (\ref{eqII5}) provide $\mathcal{L}[\sigma^2(t)](s)=2D_\alpha/s^{1-\alpha}$. From the relation $\mathcal{L}^{-1}[1/s^{1+\mu}](t)=t^\mu/\Gamma(1+\mu)$, $\mu>-1$, we obtain well--known result
\begin{equation}\label{eqIII9}
\sigma^2_\alpha(t)=\frac{2D_\alpha}{\Gamma(1+\alpha)}t^\alpha.
\end{equation}
From Eqs. (\ref{eqII12}) and (\ref{eqIII8}) we get the first passage time distribution, derived previously in \cite{barkai2001}
\begin{eqnarray}\label{eqIII10}
F_\alpha(t;x_0,x_M)=\tilde{f}_{0,\alpha/2}\left(t;\frac{x_M-x_0}{\sqrt{D_\alpha}}\right)=\frac{\alpha|x_M-x_0|}{2\sqrt{D_\alpha}t^{1+\alpha/2}}\\
\times\sum_{j=0}^\infty\frac{1}{j!\Gamma(1-\alpha(j+1)/2)}\left(-\frac{(x_M-x_0)}{\sqrt{D_\alpha t^\alpha}}\right)^j.\nonumber
\end{eqnarray}
Eq. (\ref{eqIII10}) can be derived by integrating and differentiating Eq. (\ref{eqIII8}) term by term according to the formula Eq. (\ref{eqII12}). Another method is to insert Eq. (\ref{eqIII6}) into the ordinary Laplace transform of Eq. (\ref{eqII12}) and then invert the obtained transform using Eq. (\ref{eqIII7}).
In the long time limit Eq. (\ref{eqIII10}) reads
\begin{equation}\label{eqIII11}
F_\alpha(t;x_0,x_M)=\frac{\alpha|x_M-x_0|}{2\sqrt{D_\alpha}\Gamma(1-\alpha/2)t^{1+\alpha/2}}.
\end{equation}
Comparing Eqs. (\ref{eqIII8}) and (\ref{eqIII10}) we find
\begin{equation}\label{eqIII12}
F_\alpha(t;x_0,x_M)=\frac{\alpha|x_M-x_0|}{t}P_\alpha(|x_M-x_0|,t|0).
\end{equation}

\section{G--subdiffusion equation with fractional Caputo time derivative with respect to another function\label{SecIV}}

The composite subdiffusion equation contains the fractional Caputo time derivative with respect to another function $g$, the order of the equation is $\alpha\in(0,1)$. In this section, we first show Caputo derivatives with respect to another function $g$, and then describe the $g$--subdiffusion equation.

\subsection{Fractional Caputo derivative with respect to another function\label{SecIVa}}

We show the basis of fractional calculus where the fractional Caputo derivative with respect to another function $g$ is involved; hereinafter the derivative is called $g$--Caputo fractional derivative. We assume that the function $g$, which values are given in a time unit, satisfies the conditions $g(0)=0$, $g(\infty)=\infty$, and $g'(t)>0$ for $t>0$. The following considerations are made for the case of $\alpha\in(0,1)$. 

The $g$--Caputo fractional derivative of the order $\alpha\in(0,1)$ is defined as \cite{almeida2017,fahad2021,jarad}
\begin{equation}\label{eqIVa1}
\frac{^Cd^\alpha_g f(t)}{dt^\alpha}=\frac{1}{\Gamma(1-\alpha)}\int_0^t [g(t)-g(u)]^{-\alpha}f'(u)du.
\end{equation}
When $g(t)\equiv t$, the $g$-Caputo fractional derivative takes the form of the ordinary Caputo derivative. 
The first order Caputo derivative with respect to the function $g$ is 
\begin{equation}\label{eqIVa2}
\frac{^C d_g f(t)}{dt}={\rm lim}_{\alpha\rightarrow 1^-}\frac{^Cd^\alpha_g f(t)}{dt^\alpha}=\frac{f'(t)}{g'(t)}.
\end{equation}

An effective method of solving linear equations with the $g$--Caputo time derivative is the $g$--Laplace transform method.
The $g$-Laplace transform is defined as \cite{fahad2021,jarad}
\begin{equation}\label{eqIVa3}
\mathcal{L}_g[f(t)](s)=\int_0^\infty {\rm e}^{-s g(t)}f(t)g'(t)dt.
\end{equation}
This transform has the property for $\alpha\in(0,1)$ \cite{fahad2021,jarad}
\begin{equation}\label{eqIVa4}
\mathcal{L}_g\left[\frac{^C d^\alpha_g f(t)}{dt^\alpha}\right](s)=s^\alpha\mathcal{L}_g[f(t)](s)-s^{\alpha-1}f(0).
\end{equation}
The $g$--Laplace transform is related to the ordinary Laplace transform as follows
\begin{equation}\label{eqIVa5}
\mathcal{L}_g[f(t)](s)=\mathcal{L}[f(g^{-1}(t))](s).
\end{equation}
Eq. (\ref{eqIVa5}) provides the rule
\begin{equation}\label{eqIVa6}
\mathcal{L}_g[f(t)](s)=\mathcal{L}[h(t)](s)\Leftrightarrow f(t)=h(g(t)).
\end{equation}
The above formula is helpful in calculating the inverse $g$--Laplace transform if the inverse ordinary Laplace transform is known. For example, 
\begin{equation}\label{eqIVa7}
\mathcal{L}_g^{-1}\left[\frac{1}{s^{\mu+1}}\right](t)=\frac{g^\mu(t)}{\Gamma(1+\mu)}, 
\end{equation}
$\mu>-1$, and
\begin{eqnarray}\label{eqIVa8}
\mathcal{L}_g^{-1}\left[s^\nu {\rm e}^{-as^\beta}\right](t)\equiv \tilde{f}_{\nu,\beta}(g(t);a)\\
=\frac{1}{g^{1+\nu}(t)}\sum_{j=0}^\infty \frac{1}{j!\Gamma(-\nu-\beta j)}\left(-\frac{a}{g^\beta(t)}\right)^j,\nonumber
\end{eqnarray}
where $a,\beta>0$.

\subsection{G--subdiffusion equation \label{SecIVb}}

In the following, functions related to $g$-subdiffusion are denoted by the index $g$. The $g$-subdiffusion equation is defined as \cite{kd2021a}
\begin{equation}\label{eqIVb1}
\frac{^C \partial^{\alpha}_g P_{g}(x,t|x_0)}{\partial t^\alpha}=D_\alpha\frac{\partial^2 P_{g}(x,t|x_0)}{\partial x^2},
\end{equation}
where $\alpha\in(0,1)$. As in the ordinary subdiffusion equation, the subdiffusion coefficient $D_\alpha$ is given in the units of ${\rm m^2/s^\alpha}$. When $g(t)\equiv t$, Eq. (\ref{eqIVb1}) takes the form of the ordinary subdiffusion equation. 

The $g$--subdiffusion equation can be solved by means of the $g$--Laplace transform method. Due to Eq. (\ref{eqIVa4}), in terms of the $g$-Laplace transform Eq. (\ref{eqIVb1}) reads
\begin{eqnarray}\label{eqIVb2}
s^\alpha \mathcal{L}_g[P_g(x,t|x_0)](s)-s^{\alpha-1}P_g(x,0|x_0)\\
=D_\alpha\frac{\partial^2 \mathcal{L}_g[P_g(x,t|x_0)](s)}{\partial x^2}.\nonumber
\end{eqnarray}

The calculations for solving Eq. (\ref{eqIVb1}) are similar to those for solving the ordinary subdiffusion equation using the ordinary Laplace transform method. The solution to Eq. (\ref{eqIVb2}) for the boundary conditions $\mathcal{L}_g[P_g(\pm\infty,t|x_0)](s)=0$ and the initial condition $P_g(x,0|x_0)=\delta(x-x_0)$ is 
\begin{equation}\label{eqIVb3}
\mathcal{L}_g[P_g(x,t|x_0)](s)=\frac{1}{2\sqrt{D_\alpha}s^{-1+\alpha/2}}\;{\rm e}^{-\frac{|x-x_0|}{\sqrt{D_\alpha}}s^{\alpha/2}}.
\end{equation}
From Eqs. (\ref{eqIVa8})  and (\ref{eqIVb3}) we get
\begin{eqnarray}\label{eqIVb4}
P_g(x,t|x_0)=\frac{1}{2\sqrt{D_\alpha}}\tilde{f}_{-1+\alpha/2,\alpha/2}\left(g(t);\frac{|x-x_0|}{\sqrt{D_\alpha}}\right)\\
=\frac{1}{2\sqrt{D_\alpha g^\alpha (t)}}\sum_{j=0}^\infty \frac{1}{j!\Gamma(1-\alpha(j+1)/2)}\left(-\frac{|x-x_0|}{\sqrt{D_\alpha g^\alpha(t)}}\right)^j.\nonumber
\end{eqnarray}
Using the $g$--Laplace transform of Eq. (\ref{eqII5}) and Eqs. (\ref{eqIVa7}), (\ref{eqIVb3}) we get \cite{kd2021a}
\begin{equation}\label{eqIVb5}
\sigma^2_{g}(t)=\frac{2D_\alpha}{\Gamma(1+\alpha)}g^\alpha(t).
\end{equation}

Putting Eq. (\ref{eqIVb4}) to Eq. (\ref{eqII12}) we obtain
\begin{eqnarray}\label{eqIVb6}
F_g(t;x_0,x_M)=\frac{\alpha|x_M-x_0|g'(t)}{2\sqrt{D_\alpha}g^{1+\alpha/2}(t)}\\
\times\sum_{j=0}^\infty\frac{1}{j!\Gamma(1-\alpha(j+1)/2)}\left(-\frac{|x_M-x_0|}{\sqrt{D_\alpha g^\alpha (t)}}\right)^j.\nonumber
\end{eqnarray}
In the limit of long time we have
\begin{equation}\label{eqIVb7}
F_g(t;x_0,x_M)=\frac{\alpha|x_M-x_0|}{2\sqrt{D_\alpha}\Gamma(1-\alpha/2)}\frac{g'(t)}{g^{1+\alpha/2}(t)}.
\end{equation}
Comparing Eqs. (\ref{eqIVb4}) and (\ref{eqIVb6}) we get
\begin{equation}\label{eqIVb8}
F_g(t;x_0,x_M)=\frac{\alpha|x_M-x_0|g'(t)}{g(t)}P_g(x_M,t|x_0).
\end{equation}

Normal diffusion and ordinary subdiffusion are qualitatively different. However, the normal diffusion equation can be treated as a special case of the ordinary subdiffusion equation obtained in the limit of $\alpha\rightarrow 1^-$. Thus, normal diffusion as an initial process is not excluded from the model. Eqs. (\ref{eqIVa2}) and (\ref{eqIVb1}) provide
\begin{equation}\label{eqIVb9}
\frac{\partial P(x,t|x_0)}{\partial t}=g'(t)D_1\frac{\partial^2 P(x,t|x_0)}{\partial x^2},
\end{equation}
The Green's function for the above equation can be obtained by taking the limit $\alpha\rightarrow 1^-$ in Eq. (\ref{eqIVb4}). We get $P(x,t|x_0)={\rm e}^{-\frac{(x-x_0)^2}{D_1 g(t)}}/(2\sqrt{\pi D_1 g(t)})$. This case is also not excluded from the model as a special case of $g$--subdiffusion. However, we do not call the process "$g$--normal diffusion" because it does not have the normal diffusion features for $g(t)\neq t$, for example $\sigma_g^2(t)\neq 2D_1 t$.

\section{Fractional superdiffusion\label{SecV}}

Superdiffusion is usually described by a superdiffusion equation with the fractional derivative with respect to a spatial variable
\begin{equation}\label{eqV1}
\frac{\partial P_\gamma(x,t|x_0)}{\partial t}=D_\gamma\frac{\partial^\gamma P_\gamma(x,t|x_0)}{\partial x^\gamma},
\end{equation}
here $\gamma\in (1,2)$, functions and parameters related to the fractional superdiffusion are denoted here by the index $\gamma$. The fractional derivative in Eq. (\ref{eqV1}) is often defined by its Fourier transform $\mathcal{F}[f(x)](k)=\int_{-\infty}^\infty {\rm e}^{ikx}f(x)dx$, namely 
\begin{equation}\label{eqV2}
\mathcal{F}\left[\frac{d^\gamma f(x)}{dx^\gamma}\right](k)=-|k|^\gamma\mathcal{F}[f(x)](k).
\end{equation} 
Various representations of this derivative have been discussed, as Riesz, Riesz--Weyl, and Riesz--Feller derivatives, see \cite{mk,compte,saichev1997,gorenflo1998,saxena2014,mainardi2001}; for the sake of simplicity, we call the derivative defined by Eq. (\ref{eqV2}) the fractional derivative of Riesz type. We add that the explicit form of the fractional spatial derivative is not necessary here since the relation (\ref{eqV2}) is sufficient for the determination of Green's function for Eq. (\ref{eqV1}), see Appendix. The Green's function is
\begin{eqnarray}\label{eqV3}
 P_\gamma(x,t|x_0)=\frac{1}{\sqrt{\pi}|x-x_0|}\\
\times H^{1 1}_{1 2}\left(\frac{|x-x_0|^\gamma}{2^\gamma D_\gamma t}\left| 
			\begin{array}{cc}
        (1,1) \\
        (1/2,\gamma/2)\;(1,\gamma/2)
      \end{array}\right. \right)\;,\nonumber
\end{eqnarray}
where $H$ denotes the H--Fox function; the method of solving Eq. (\ref{eqV1}) and the series representation of the H--Fox function are presented in Appendix. Properties of the H--Fox functions and their applications in modeling anomalous diffusion processes can be found, among others, in Ref. \cite{wyss1986,schneider1989,mainardi2005}. Due to Eq. (\ref{a1}) in Appendix the series representation of the Green's function Eq. (\ref{eqV3}) reads
\begin{eqnarray}\label{eqV4}
P_\gamma(x,t|x_0)=\frac{1}{\gamma\sqrt{\pi}(D_\gamma t)^{1/\gamma}}\\
\times\sum_{j=0}^\infty \frac{\Gamma(1/\gamma+2j/\gamma)}{j!\Gamma(1/2+j)}\left(-\frac{(x-x_0)^2}{4(D_\gamma t)^{2/\gamma}}\right)^j.\nonumber
\end{eqnarray}
The domain in which the above series is convergent depends on the parameters $\gamma$ and $D_\gamma$. Analysis of plots of the function Eq. (\ref{eqV4}) making for different values of the parameters shows that the series is convergent for $\gamma\in(1,2)$ when $(x-x_0)^2/[4(D_\gamma t)^{2/\gamma}]<\mu$ with (approximately) $\mu=12(\gamma-0.9)^2$. For $\gamma=2$ Eqs. (\ref{eqV4}) takes the form of Green's function for normal diffusion $P_{\gamma=2}(x,t|x_0)={\rm e}^{-\frac{(x-x_0)^2}{4D_\gamma t}}/(2\sqrt{\pi D_\gamma t})$, $x\in(-\infty,\infty)$. 

The Green's function for Eq. (\ref{eqV1}) can be expressed in a form other than Eq. (\ref{eqV4}). For $x_0=0$, in Ref. \cite{mk} this function is (see also \cite{west,jespersen}), here $P_\gamma(x,t)\equiv P_\gamma(x,t|0)$,
\begin{eqnarray}\label{eqV3a}
 P_\gamma(x,t)=\frac{1}{\gamma|x|}
H^{1 1}_{2 2}\left(\frac{|x|}{(D_\gamma t)^{1/\gamma}}\left| 
			\begin{array}{cc}
        (1,1/\gamma)\;(1.1/2) \\
        (1,1)\;(1,1/2)
      \end{array}\right. \right)\;,
\end{eqnarray}
and in Refs. \cite{mainardi2001,mainardi2005}, in which the case of $D_\gamma=1$ has been considered, this function takes the following form 
\begin{eqnarray}\label{eqV3b}
 P_\gamma(x,t)=\frac{1}{\gamma t^{1/\gamma}}
H^{1 1}_{2 2}\left(\frac{|x|}{t^{1/\gamma}}\left| 
			\begin{array}{cc}
        (1-1/\gamma,1/\gamma)\;(1/2,1/2) \\
        (0,1)\;(1/2,1/2)
      \end{array}\right. \right)\;.
\end{eqnarray}
The Green's functions Eqs. (\ref{eqV3a}) and (\ref{eqV3b}) can be expressed as a series Eq. (\ref{eqV4}) ($D_\gamma=1$ in the latter function), the proof is in Appendix. Thus, despite the different forms, the Green's functions (\ref{eqV3}), (\ref{eqV3a}), and (\ref{eqV3b}) are equivalent to each other.

From Eqs. (\ref{eqII5}) and (\ref{eqV4}) we get
\begin{equation}\label{eqV8}
\sigma_\gamma^2(t)=\frac{8D^{2/\gamma}\kappa}{\gamma\sqrt{\pi}}t^{2/\gamma},
\end{equation}
where 
\begin{equation}\label{eqV9}
	\kappa=\int_0^\infty u^{-1+2/\gamma} H^{1 1}_{1 2}\left(u\left| 
			\begin{array}{cc}
        (1,1) \\
        (1/2,\gamma/2)\;(1,\gamma/2)
      \end{array}\right. \right)du\;.
	\end{equation}
Since $H^{1 1}_{1 2}\left(u\left| 
			\begin{array}{cc}
        (1,1) \\
        (1/2,\gamma/2)\;(1,\gamma/2)
      \end{array}\right. \right)\sim \mathcal{O}(1)$ when $u\rightarrow\infty$ (see Eq. (\ref{a3}) in Appendix), $\kappa=\infty$, and  consequently $\sigma_\gamma^2(t)=\infty$ for $t>0$. The problem of the divergence of MSD generated by the Green function being a solution to Eq. (\ref{eqV1}) has been discussed in \cite{mk,west}. In the next section, we consider a model that describes superdiffusion in the long time limit and leads to a finite MSD. 
			
Instead of $\sigma_\gamma^2$, we use the function $F_\gamma$ to characterize fractional superdiffusion.
From Eqs. (\ref{eqII12}) and (\ref{eqV4}) we obtain
\begin{eqnarray}\label{eqV5}
F_\gamma(t;x_0,x_M)=\frac{2|x_M-x_0|}{\gamma^2\sqrt{\pi}(D_\gamma)^{1/\gamma}t^{1+1/\gamma}}\\
\times\sum_{j=0}^\infty \frac{\Gamma(1/\gamma+2j/\gamma)}{j!\Gamma(1/2+j)}\left(-\frac{(x_M-x_0)^2}{4(D_\gamma t)^{2/\gamma}}\right)^j.\nonumber
\end{eqnarray}
In the limit of long time there is
\begin{equation}\label{eqV6}
F_\gamma(t;x_0,x_M)=\frac{2|x_M-x_0|\Gamma(1/\gamma)}{\pi\gamma^2D_\gamma^{1/\gamma}}\frac{1}{t^{1+1/\gamma}}.
\end{equation}
Comparing Eqs. (\ref{eqV4}) and (\ref{eqV5}) we find
\begin{eqnarray}\label{eqV7}
F_\gamma(t;x_0,x_M)=\frac{2|x_M-x_0|}{\gamma t}P_\gamma(x_M,t|x_0).
\end{eqnarray}

\section{From ordinary subdiffusion to superdiffusion\label{SecVI}}

We propose a model for a smooth transition from ordinary subdiffusion with parameters $\alpha\in(0,1)$ and $D_\alpha$ to fractional superdiffusion with parameters $\gamma\in(1,2)$ and $D_\gamma$. We use the $g$--subdiffusion equation to describe the transition process. Since the $g$--subdiffusion and fractional superdiffusion equations unequivocally determine the Green's functions $P_g$ and $P_\gamma$ respectively, the following considerations are based on these functions. 

In some initial time interval, the process is described by the ordinary subdiffusion equation Eq. (\ref{eqIII1}). In the limit of long time the process is described by the fractional superdiffusion equation Eq. (\ref{eqV1}). This assumption is equivalent to that for very short time the Green's function $P_g$ is given by Eq. (\ref{eqIII8}) and for long time is expressed by Eq. (\ref{eqV4}). The smooth transition means that the first derivative of the Green's function describing this process exists and is continuous for $x\in(-\infty,\infty)$ and $t>0$. 

\begin{figure}[htb]
\centering{%
\includegraphics[scale=0.4]{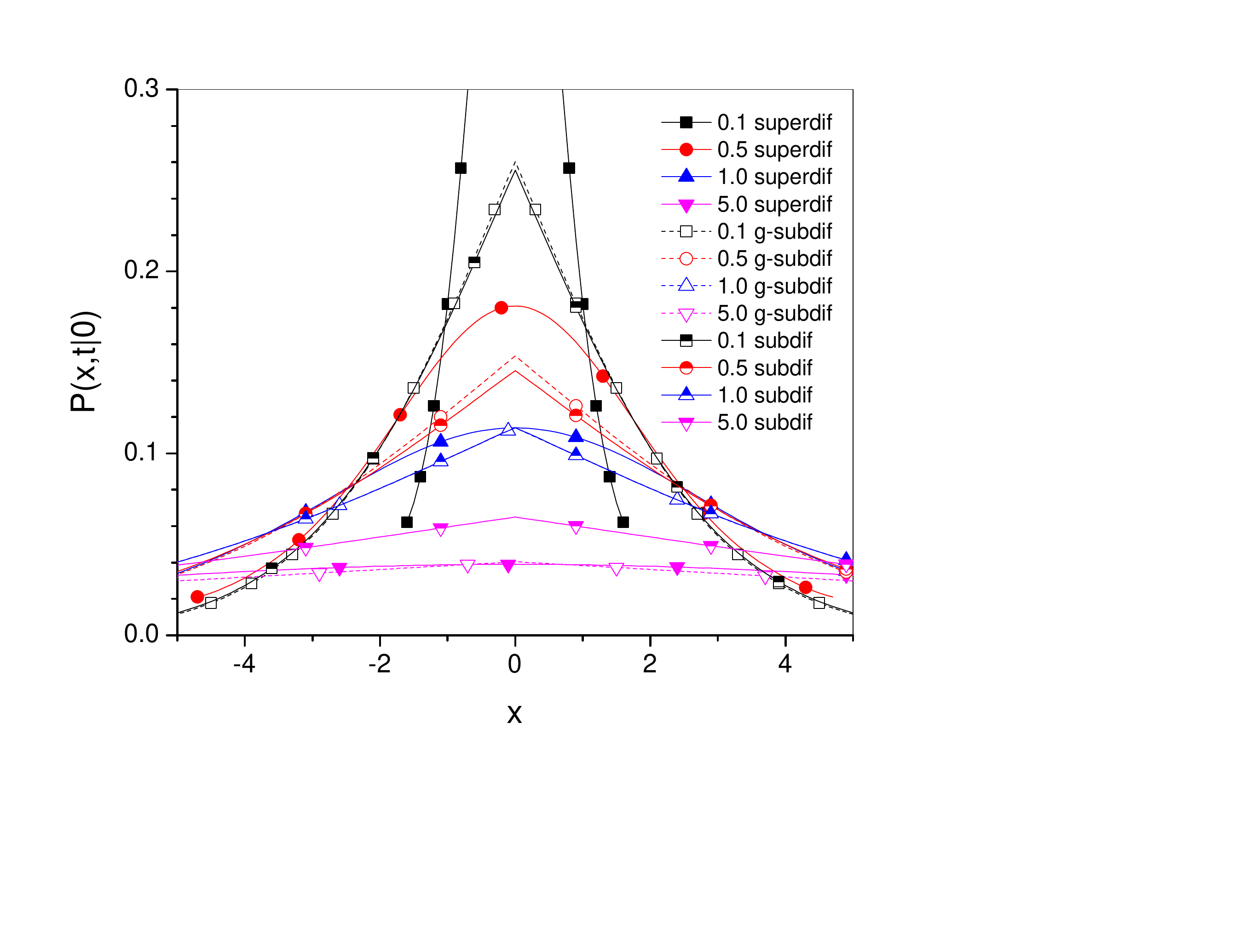}}
\caption{Plots of Green's functions for fractional superdiffusion Eq. (\ref{eqV4}) (solid lines with filled symbols) with $\gamma=1.5$ and $D_\gamma=4$, for ordinary subdiffusion (solid lines with half--filled symbols) Eq. (\ref{eqIII8}) with $\alpha=0.7$ and $D_\alpha=10$, and for $g$--subdiffusion Eq. (\ref{eqIVb4}) (dashed lines with open symbols) with $g$ Eq. (\ref{eqVI7}) and $E$ Eq. (\ref{eqVI4}), $A=1$, $\nu=1.2$, and $\alpha$, $D_\alpha$ given above. The plots are made for times given in the legend. The plot of Green's function for fractional superdiffusion for $t=0.1$ (solid lines with filled squares) is presented in the interval $x\in(-1.6,1.6)$ only due to limited domain in which the series occurring in Eq. (\ref{eqV4}) is convergent.}
\label{fig1}
\end{figure}

\begin{figure}[htb]
\centering{%
\includegraphics[scale=0.4]{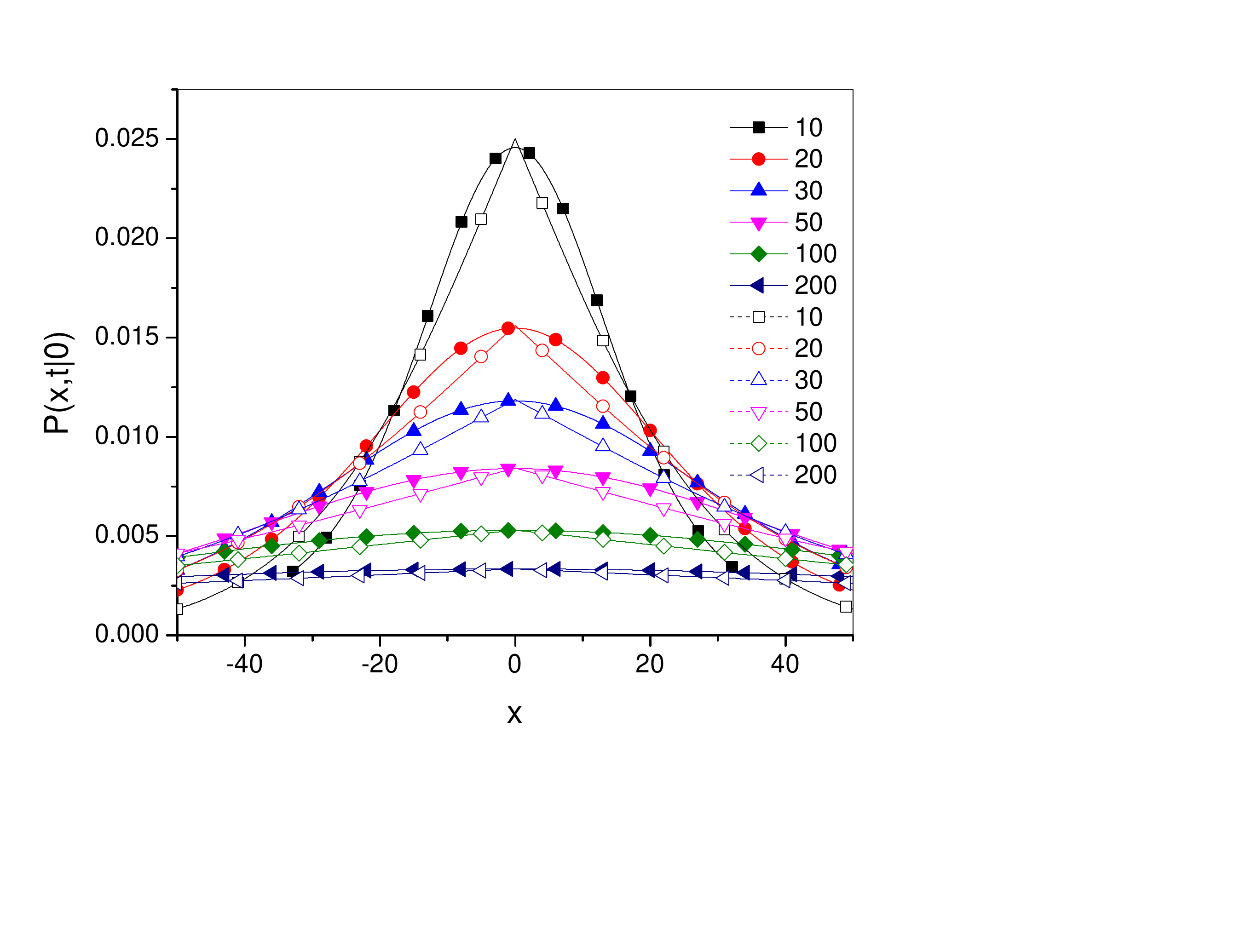}}
\caption{Plots of Green's functions for fractional superdiffusion equation (\ref{eqV4}) (solid lines with filled symbols), and for $g$--subdiffusion equation (\ref{eqIVb4}) (dashed lines with open symbols) for times given in the legend, the values of the parameters are the same as in the caption of Fig. \ref{fig1}.}
\label{fig2}
\end{figure}

\begin{figure}[htb]
\centering{%
\includegraphics[scale=0.4]{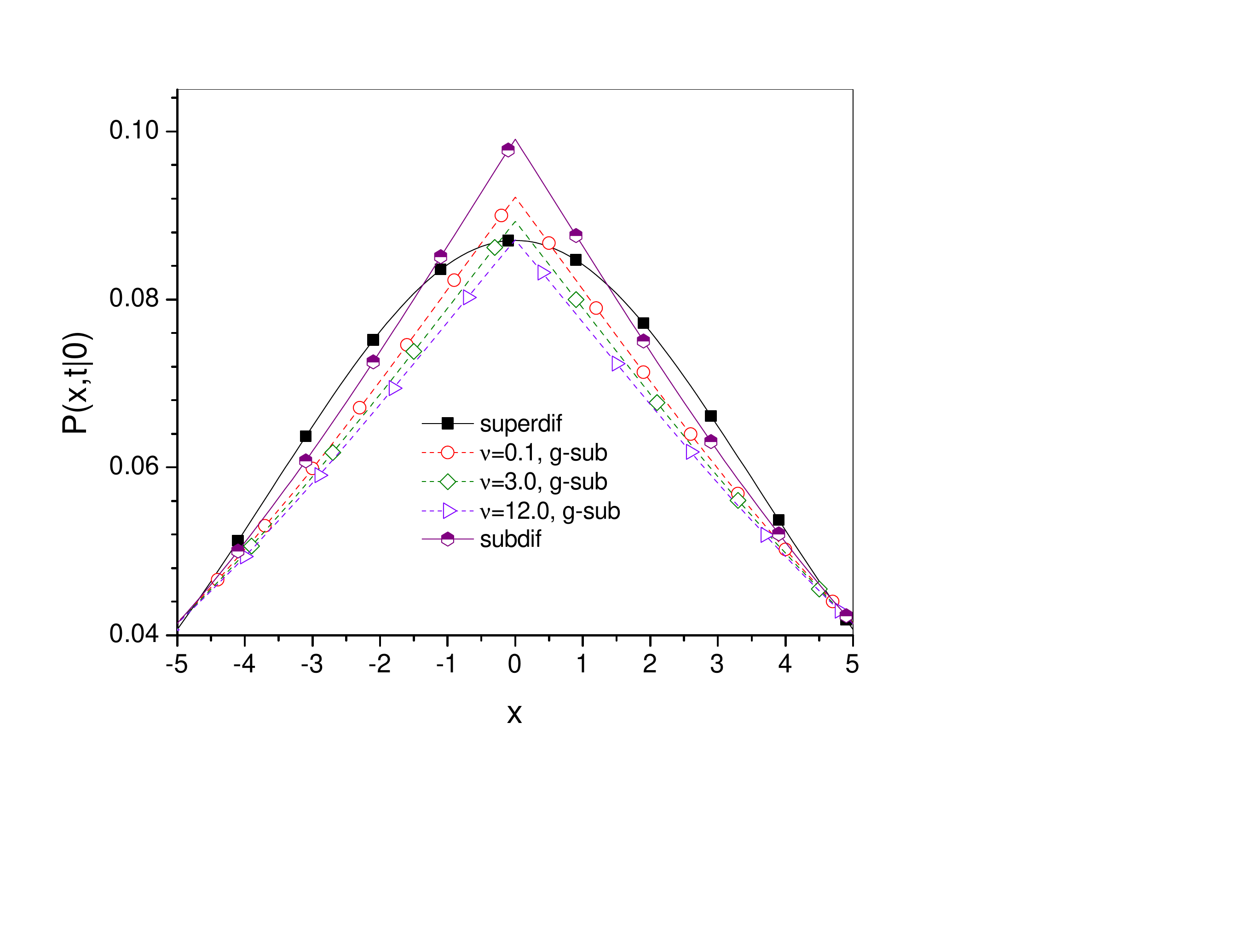}}
\caption{Plots of Green's functions for fractional superdiffusion Eq. (\ref{eqV4}) (solid lines with filled symbols), for $g$--subdiffusion Eq. (\ref{eqIVb4}) (dashed lines with open symbols) for $\nu$ given in the legend, and for ordinary subdiffusion (solid lines with half--filled symbols) Eq. (\ref{eqIII8}), $t=1.5$, the values of the other parameters are the same as in the caption of Fig. \ref{fig1}.}
\label{fig3}
\end{figure}

We propose the function $g$ that makes the Green's function Eq. (\ref{eqIVb4}) have the properties mentioned above. First, we determine $g$ when $t\rightarrow 0$ and $t\rightarrow\infty$. Then, we propose the function $g$ over the entire time domain.
  
In the limit of short time $P_g$ Eq. (\ref{eqIVb4}) takes the form of the Green's function for ordinary subdiffusion $P_\alpha$ Eq. (\ref{eqIII8}) with parameters $\alpha$ and $D_\alpha$. This condition is met when 
\begin{equation}\label{eqVI1}
g(t)=t,\;t\rightarrow 0.
\end{equation}
We note that Eqs. (\ref{eqIVb8}) and (\ref{eqV7}) provide the condition that $F_g(t;x_0,x)\equiv F_\gamma(t;x_0,x)$ and $P_g(x,x,t|x_0)\equiv P_\gamma(x,t|x_0)$ are satisfied simultaneously when $g'(t)/g(t)=2/(\gamma\alpha)$. The solution to the above equation is 
\begin{equation}\label{eqVI2}
g(t)=Et^{\frac{2}{\gamma\alpha}}.
\end{equation} 
where $E$ is a positive coefficient. Thus, if the function $g$ is given by Eq. (\ref{eqVI2}), then
\begin{equation}\label{eqVI3} 
F_g(t;x_0,x)\equiv F_\gamma(t;x_0,x)\Leftrightarrow P_g(x,t|x_0)\equiv P_\gamma(x,t|x_0).
\end{equation} 

The Green's functions for $g$--subdiffusion Eq. (\ref{eqIVb4}) and for fractional superdiffusion Eq. (\ref{eqV4}) are qualitatively different. They can be equivalent in the long time limit only. According to the statement, the condition $P_g(x,t|x_0)\equiv P_\gamma(x,t|x_0)$ when $t\rightarrow\infty$ is met if the functions Eqs. (\ref{eqIVb7}) and (\ref{eqV6}) are the same. Equality of these functions provides
\begin{equation}\label{eqVI4}
E=\left(\frac{\pi\gamma D_\gamma^{1/\gamma}}{2\sqrt{D_\alpha}\Gamma(1/\gamma)\Gamma(1-\alpha/2)}\right)^{2/\alpha}.
\end{equation}
Thus, we assume that in the long time limit the function $g$ is given by Eq. (\ref{eqVI2}) with the coefficient $E$ Eq. (\ref{eqVI4}).

Guided by Eqs. (\ref{eqVI1}) and (\ref{eqVI2}), the latter for $t\rightarrow\infty$, we define the function $g$ as
\begin{equation}\label{eqVI5}
g(t)=a(t)t+[1-a(t)]Et^{\frac{2}{\gamma\alpha}},
\end{equation}
where the non--negative function $a$ fulfils the conditions $a(0)=1$, $a(\infty)=0$, and ensures that $g'(t)>0$, $t>0$. 

In the following, we consider the function
\begin{equation}\label{eqVI6}
a(t)=\frac{1}{1+At^\nu},
\end{equation}
where $A$ and $\nu$ are positive parameters. 
Then, 
\begin{equation}\label{eqVI7}
g(t)=\frac{t+AEt^{\frac{2}{\gamma\alpha}+\nu}}{1+At^\nu}.
\end{equation}

In Figs. \ref{fig1} and \ref{fig2} the Green's functions for ordinary subdiffusion $P_\alpha$, fractional superdiffusion $P_\gamma$, and $g$--subdiffusion $P_g$ are presented for different times, the values of all parameters are given in arbitrarily chosen units. The function $P_g$ is close to $P_\alpha$ for small times and approaches $P_\gamma$ with time. With the assumed values of parameters, for $t>10$ the plots of $P_g$ and $P_\gamma$ are very close to each other. In Fig. \ref{fig3} plots of $P_g$ are made for different values of $\nu$. For larger values of this parameter $P_g$ moves away from $P_\alpha$ faster.

\section{Scaling properties\label{SecVII}}

In the scaling method, the following variable transform $(x,t)\rightarrow(x',t')$ in an equation, its solution, and boundary and initial conditions is done,
\begin{equation}\label{eqVII1}
x'=b^\mu x\;,\;t'=b^\nu t,
\end{equation}
where $b$ is a positive parameter.
If a solution to the equation is transformed according to the formula 
\begin{equation}\label{eqVII2}
P(x',t'|x'_0)=b^\lambda P(x,t|x_0),
\end{equation}
the general form of the solution is
\begin{equation}\label{eqVII3}
P(x,t|x_0)=t^{\lambda/\nu}\Phi(\eta),
\end{equation}
where $\eta\sim x/t^{\mu/\nu}$ is a variable invariant with respect to the transform Eq. (\ref{eqVII1}). Let $v=t'/t$, then $x'=v^{\mu/\nu}x$ and $P(x',t'|x'_0)=v^{\lambda/\nu}P(x,t|x_0)$. The function $P(x,t|x_0)$ has a scaling property when the plots composed of points $(x',P(x',t'|x'_0))\equiv(v^{\mu/\nu}x,v^{\lambda/\nu}P(x,t|x_0))$ are identical to those of $(x,P(x,t|x_0))$ for $t>0$ and all $x$ from the domain of the function.

For fractional superdiffusion we have, see Eq. (\ref{eqV4}),
\begin{equation}\label{eqVII4}
P_\gamma(x,t|x_0)=t^{-1/\gamma}\Phi_\gamma(\eta_\gamma),
\end{equation}
where
\begin{equation}\label{eqVII5}
\eta_\gamma=\frac{|x-x_0|}{2(D_\gamma t)^{1/\gamma}},
\end{equation}
and
\begin{equation}\label{eqVII6}
\Phi_\gamma(\eta_\gamma)=\frac{1}{\gamma\sqrt{\pi}D_\gamma^{1/\gamma}}\sum_{j=0}^\infty \frac{\Gamma(1/\gamma+2j/\gamma)}{\Gamma(j+1/2)}\left(-\eta_\gamma^2\right)^j.
\end{equation}
Thus, for fractional superdiffusion we get $\mu/\nu=1/\gamma$ and $\lambda/\nu=-1/\gamma$.
Scaling of Green's functions for fractional superdiffusion is discussed, among others, in \cite{mainardi2001}.

\begin{figure}[htb]
\centering{%
\includegraphics[scale=0.4]{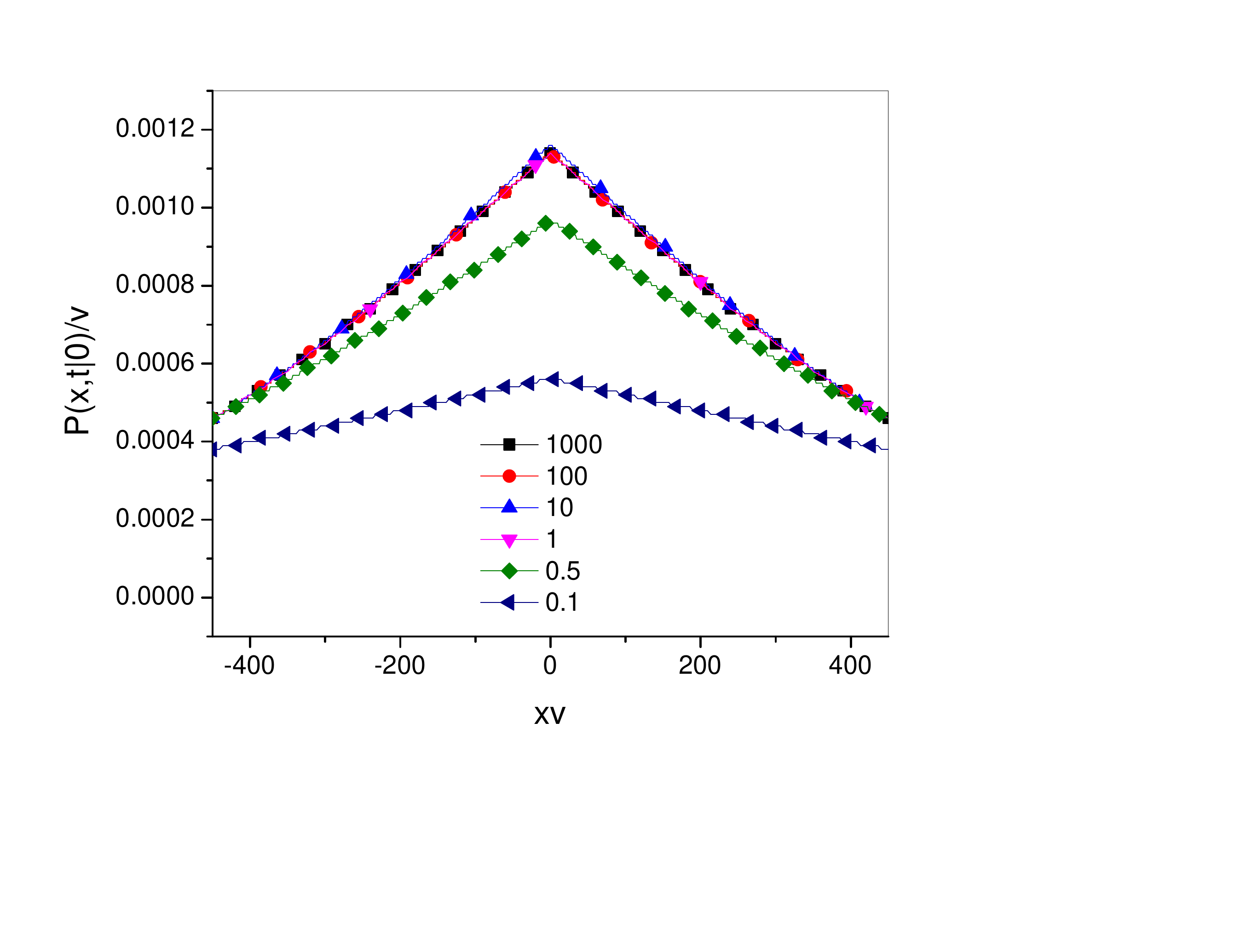}}
\caption{Scaling property of Green's functions for $g$--subdiffusion Eq. (\ref{eqIVb4}), $v=(t'/t)^{1/\gamma}$, $t'=1000$, values of $t$ are given in the legend, the other parameters are the same as in the caption of Fig. \ref{fig1}, more detailed description is in the text.}
\label{fig4}
\end{figure}

In the limit of long time, when $a(t)\approx 0$, we have $g(t)\approx Et^{2/\alpha\gamma}$. Then, from Eq. (\ref{eqIVb4}) we get
\begin{equation}\label{eqVII7}
P_g(x,t|x_0)\approx t^{-1/\gamma}\Phi_g(\eta_g),
\end{equation}
where
\begin{equation}\label{eqVII8}
\eta_g=\frac{|x-x_0|}{\sqrt{D_\alpha E^\alpha} t^{1/\gamma}},
\end{equation}
and 
\begin{equation}\label{eqVII9}
\Phi_g(\eta_g)=\frac{1}{\sqrt{2D_\alpha E^\alpha}}\sum_{j=0}^\infty \frac{1}{\Gamma(1-(j+1)\alpha/2)}(-\eta_g)^j.
\end{equation}

There is $P_g(x,t|x_0)\approx P_\gamma(x,t|x_0)$ when $t\rightarrow\infty$. In order to check whether the function $P_g$ can be considered as describing superdiffusion for a finite time we use the scaling method. We assume that the function $P_g$ describes superdiffusion when its scaling properties are like for $P_\gamma$. In practice, $P_g$ has a scaling property like $P_\gamma$ when the plots composed of points $(x,P_g(x,g(t)|x_0))$ and $(v^{1/\gamma}x,v^{-1/\gamma}P_g(x,t|x_0))$ coincide. An example of scaling the function $P_g$ is presented in Fig. \ref{fig4}; $P_g$ has the scaling property characteristic of fractional superdiffusion for $t>10$.

\section{Interpretation\label{SecVIII}}

\begin{figure}[htb]
\centering{%
\includegraphics[scale=0.4]{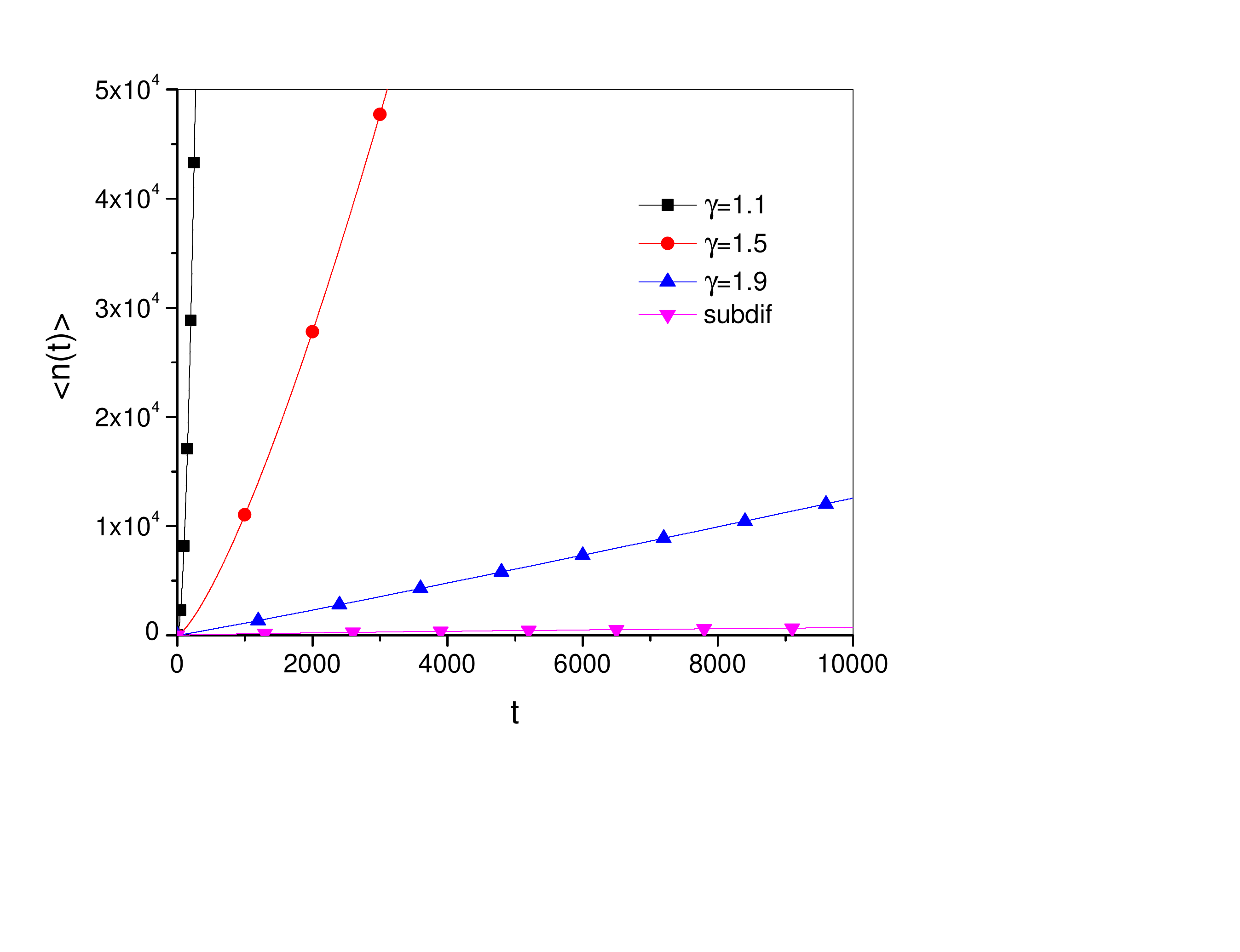}}
\caption{Time evolution of the mean number of particle steps for ordinary subdiffusion (denoted as ``subdif'' in the legend) and $g$--subdiffusion for different values of $\gamma$ given in the legend, $\tau=1$ and $\nu=1.2$.}
\label{fig5}
\end{figure}

\begin{figure}[htb]
\centering{%
\includegraphics[scale=0.4]{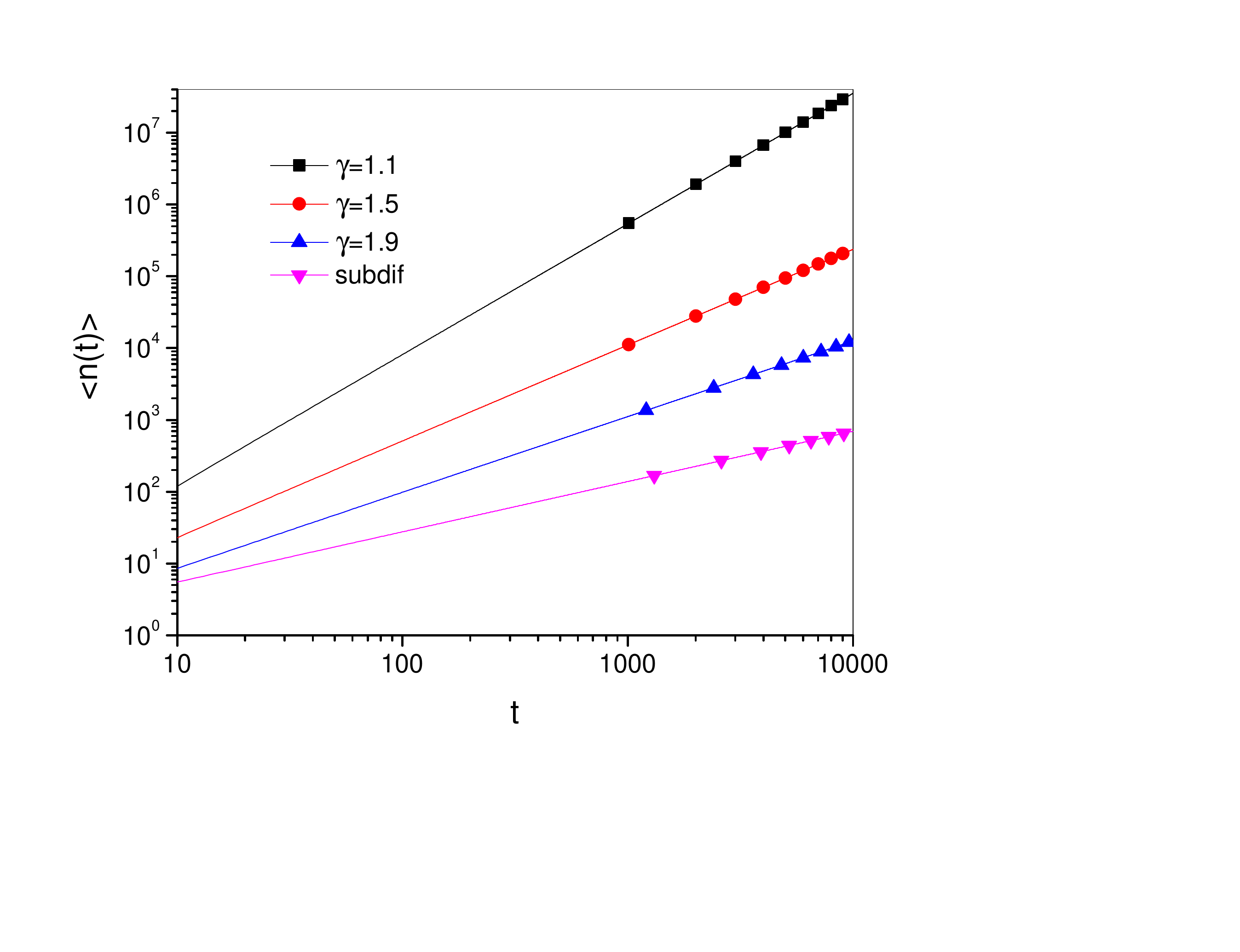}}
\caption{Time evolution of the mean number of particle steps done on a logarithmic scale, the description is similar to that in Fig. \ref{fig5}.}
\label{fig6}
\end{figure}

\begin{figure}[htb]
\centering{%
\includegraphics[scale=0.4]{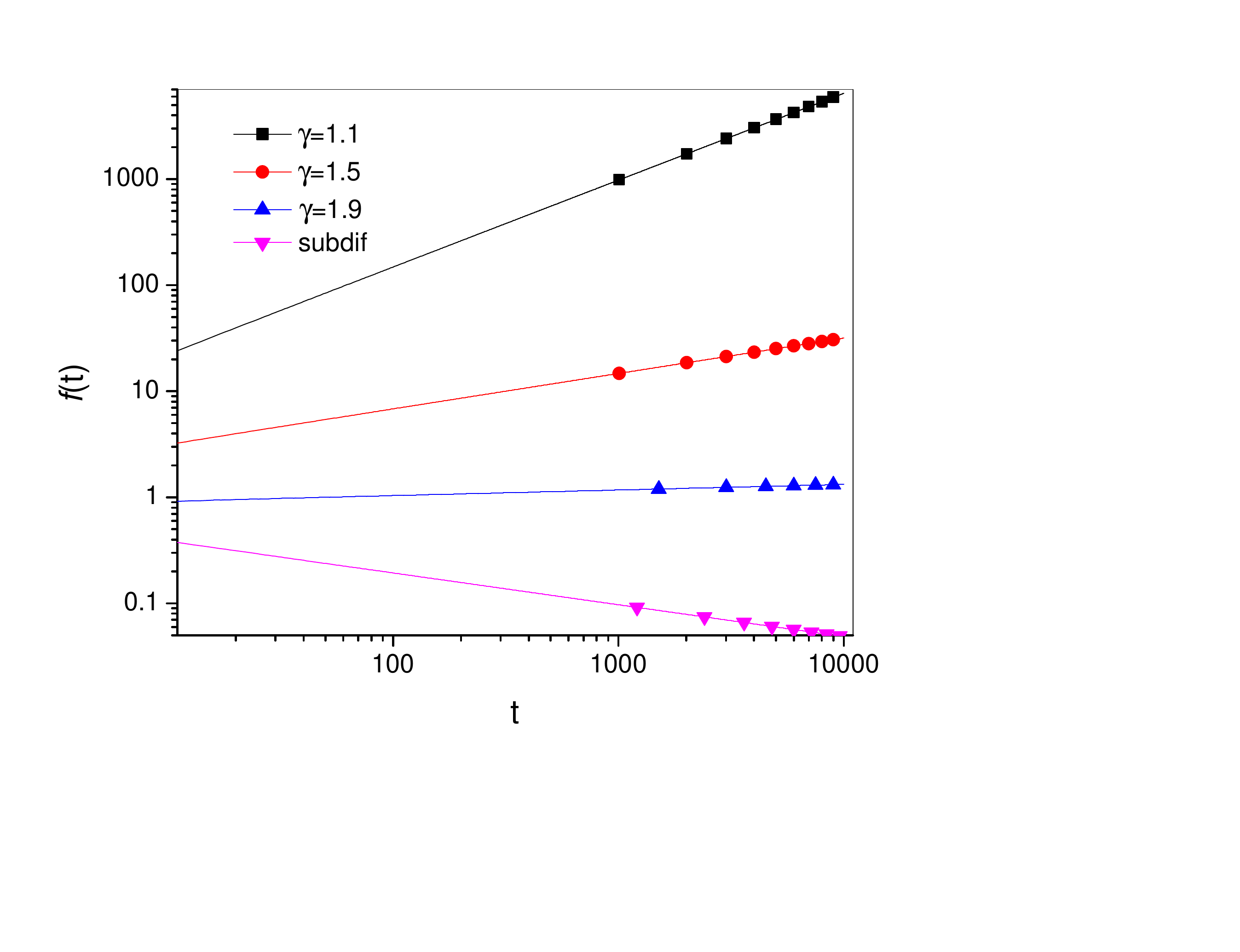}}
\caption{Time evolution of the jump frequency $f(t)=d\left\langle n(t)\right\rangle/dt$ for the cases presented in Fig. \ref{fig5}. For ordinary subdiffusion $f$ is an decreasing function of time, the superdiffusion effect is achieved when $f$ is a function increasing with time.}
\label{fig7}
\end{figure}

\begin{figure}[htb]
\centering{%
\includegraphics[scale=0.4]{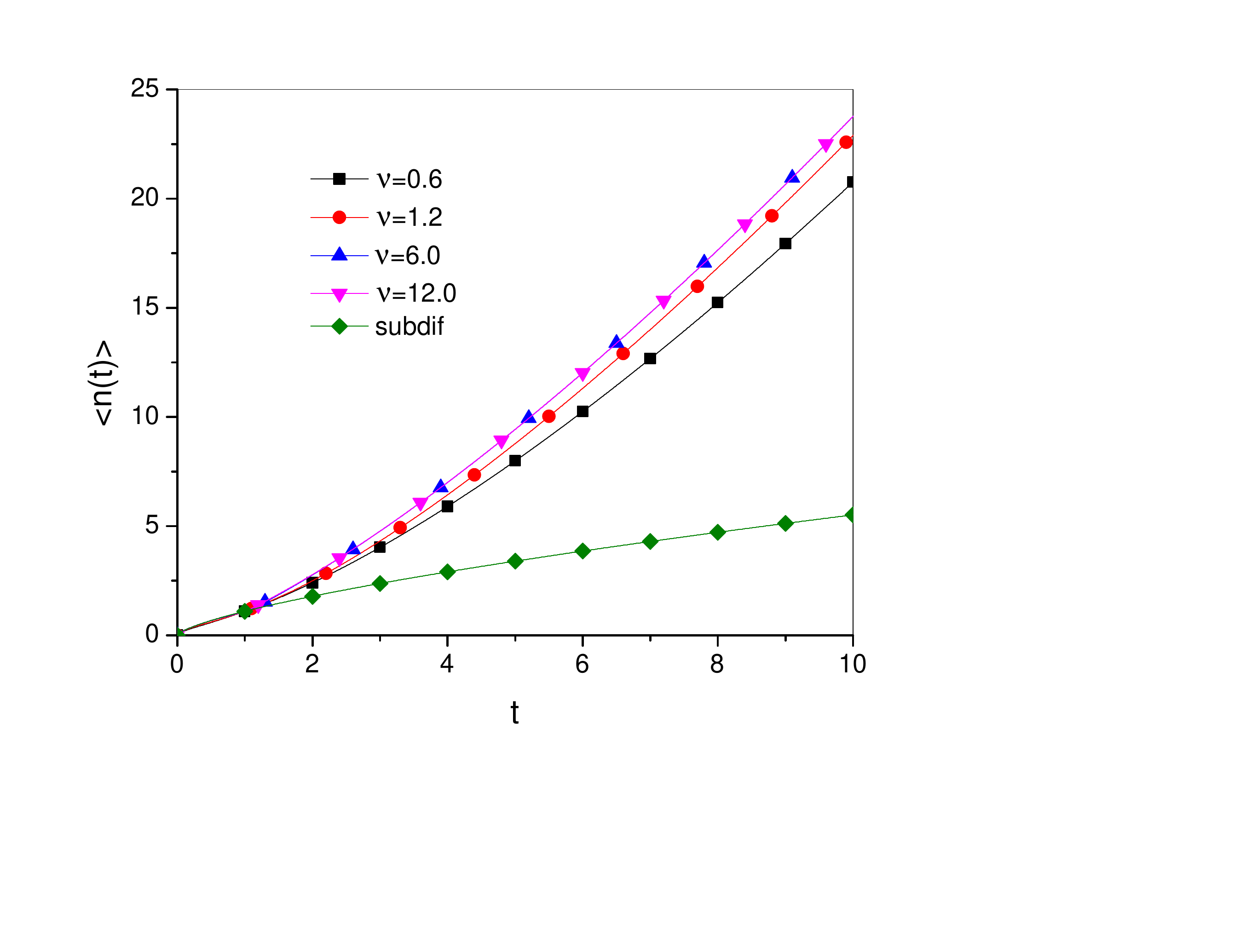}}
\caption{Time evolution of the mean number of steps for ordinary subdiffusion and $g$--subdiffusion for different values of $\nu$ given in the legend, the values of other parameters are the same as in Figs. \ref{fig1} and \ref{fig5}.}
\label{fig8}
\end{figure}

As mentioned in Sec. \ref{SecI}, the ordinary subdiffusion equation can be derived from the CTRW model, see for example Refs. \cite{mk,ks,mks,compte}. Within this model, the Green's function for ordinary subdiffusion can be considered as $P_\alpha(x,t|x_0)=\sum_{n=0}^\infty Q_{\alpha,n}(t)P_n(x|x_0)$ \cite{montroll1965,weiss1994}, where $P_n(x|x_0)$ is the probability density that diffusing particle achieves the point $x$ after $n$ steps, $x_0$ is the initial particle location, $Q_n(t)$ is the probability that the particle takes $n$ steps in the time interval $(0,t)$. There is 
$Q_{\alpha,n}(t)=(\underbrace{\psi_\alpha*\psi_\alpha*\ldots*\psi_\alpha}_{n\;times}*U_\alpha)(t)$, 
where $\psi_\alpha$ is the probability density that the particle jumps at time $t$ since its last stop, $*$ denotes the ordinary convolution of functions, $(f*h)(t)=\int_0^t f(u)h(t-u)du$, $U_\alpha(t)=1-\int_0^t\psi_\alpha(t')dt'$ is the probability that the particle does not make a jump by time $t$. The function $\psi_\alpha$ is assumed to be decreasing. Due to the following property $\mathcal{L}[(f*h)(t)](s)=\mathcal{L}[f(t)](s)\mathcal{L}[h(t)](s)$ there is $\mathcal{L}[Q_{\alpha,n}(t)](s)=\mathcal{L}^n[\psi_\alpha(t)](s)\mathcal{L}[U_\alpha(t)](s)$, where $\mathcal{L}[U_\alpha(t)](s)=(1-\mathcal{L}[\psi_\alpha(t)](s))/s$.
Assuming that the mean length of a single particle jump $\epsilon$ is finite, the function $P_\alpha$ is the Green's function for Eq. (\ref{eqIII1}) if
\begin{equation}\label{eqVIII1}
\mathcal{L}[\psi_\alpha(t)](s)=\frac{1}{1+\tau s^\alpha},
\end{equation}
where $\tau=\epsilon^2/(2D_\alpha)$, then we have
\begin{equation}\label{eqVIII2}
\mathcal{L}[U_\alpha(t)](s)=\frac{\tau s^{\alpha-1}}{1+\tau s^\alpha},
\end{equation}
see Ref. \cite{kd2021b}, and 
\begin{equation}\label{eqVIII3}
\mathcal{L}[Q_{\alpha,n}(t)](s)=\frac{\tau s^{\alpha-1}}{(1+\tau s^\alpha)^{n+1}}. 
\end{equation}
In terms of the ordinary Laplace transform the mean number of jumps $\left\langle n_{\alpha}(t)\right\rangle=\sum_{n=1}^\infty nQ_{\alpha,n}(t)$ is $\mathcal{L}[\left\langle n_{\alpha}(t)\right\rangle](s)=\mathcal{L}[\psi_\alpha(t)](s)/[s(1-\mathcal{L}[\psi_\alpha(t)](s))]$, see Ref. \cite{ks}. From the above equations we obtain
\begin{equation}\label{eqVIII4}
\mathcal{L}[\left\langle n_\alpha(t)\right\rangle](s)=\frac{1}{\tau s^{1+\alpha}},
\end{equation}
which gives
\begin{equation}\label{eqVIII5}
\left\langle n_\alpha(t)\right\rangle=\frac{t^\alpha}{\tau\Gamma(1+\alpha)},
\end{equation}
and the time evolution of jumps frequency $f_\alpha(t)=d\left\langle n_\alpha(t)\right\rangle/dt$ is
\begin{equation}\label{eqVIII6}
f_{\alpha}(t)=\frac{1}{\tau\Gamma(\alpha)t^{1-\alpha}}.
\end{equation}

The $g$--subdiffusion equation Eq. (\ref{eqIVb1}) as well as the Green's function Eq. (\ref{eqIVb4}) can be derived from a modified CTRW model (the $g$--CTRW model) \cite{kd2021b}. This model is similar to the ordinary CTRW model briefly described above. The most important differences between the models are that the modified model uses the $g$--Laplace transform and the $g$--convolution of functions instead of the ordinary Laplace transform and ordinary convolution, respectively. The $g$--convolution $f*_g h$ is defined as
\begin{eqnarray}\label{eqVIII7}
(f*_g h)(t)=[(f\circ g^{-1})*(h\circ g^{-1})](g(t))\\
=\int_0^{g(t)} f[g^{-1}(u)]h[g^{-1}(g(t)-u)]du.\nonumber
\end{eqnarray}
The introduction of $g$--convolution into the considerations is motivated by the following property which makes the method of Green's function derivation similar to that of ordinary subdiffusion \cite{jarad}
\begin{equation}\label{eqVIII8}
\mathcal{L}_g[(f*_g h)(t)](s)=\mathcal{L}_g[f(t)](s)\mathcal{L}_g[h(t)](s).
\end{equation}
We introduce the functions $\psi_g$ and $U_g$, and we assume that $P_g(x,t|x_0)=\sum_{n=0}^\infty Q_{g,n}(t)P_n(x|x_0)$, where 
\begin{equation}\label{eqVIII8a}
Q_{g,n}(t)=(\underbrace{\psi_g*\psi_g*\ldots*\psi_g}_{n\;times}*U_g)(t),
\end{equation} 
the parameters $\alpha$, $D_\alpha$, and $\tau$ are assumed to be the same as for ordinary subdiffusion. The aim is to determine relations of $\psi_g$ and $U_g$ with $\psi_\alpha$ and $U_\alpha$, respectively.
From Eqs. (\ref{eqVIII8}) and (\ref{eqVIII8a}) we get 
\begin{equation}\label{eqVIII9}
\mathcal{L}_g[Q_{g,n}(t)](s)=\mathcal{L}_g^n[\psi_g(t)](s)\mathcal{L}_g[U_g(t)](s). 
\end{equation}
As it is shown in Ref. \cite{kd2021b}, the above equations provide the Green's function for $g$--subdiffusion equation only when
\begin{equation}\label{eqVIII10}
\mathcal{L}_g[\psi_g(t)](s)=\frac{1}{1+\tau s^\alpha},
\end{equation}
and
\begin{equation}\label{eqVIII11}
\mathcal{L}_g[U_g(t)](s)=\frac{\tau s^{\alpha-1}}{1+\tau s^\alpha}.
\end{equation}
From Eqs. (\ref{eqVIII8})--(\ref{eqVIII11}) we obtain
\begin{equation}\label{eqVIII12}
\mathcal{L}_g[Q_{g,n}(t)](s)=\frac{\tau s^{\alpha-1}}{(1+\tau s^\alpha)^{n+1}}.
\end{equation}
Comparing Eqs. (\ref{eqVIII1}), (\ref{eqVIII2}), and (\ref{eqVIII3}) with Eqs. (\ref{eqVIII10}), (\ref{eqVIII11}), and (\ref{eqVIII12}), respectively, we get $\mathcal{L}_g[\psi_g(t)](s)=\mathcal{L}[\psi_\alpha(t)](s)$, $\mathcal{L}_g[U_g(t)](s)=\mathcal{L}[U_\alpha(t)](s)$, and $\mathcal{L}_g[Q_{g,n}(t)](s)=\mathcal{L}[Q_{\alpha,n}(t)](s)$. From the above equations and Eq. (\ref{eqIVa6}) we have $U_g(t)=U_\alpha(g(t))$, and
\begin{equation}\label{eqVIII13}
\psi_g(t)=\psi_\alpha(g(t)),
\end{equation}
\begin{equation}\label{eqVIII14}
Q_{g,n}(t)=Q_{\alpha,n}(g(t)).
\end{equation}
In the process considered in this paper we have $g(t)>t$, and consequently $g^{-1}(t)<t$ for $t>0$.
Eq. (\ref{eqVIII13}) shows that the probability density that a particle takes its jump at time $t$ for $g$--subdiffusion is the same as at time $g(t)$ for ordinary subdiffusion. This means that the waiting times for a jump in the $g$--subdiffusion process is typically shorter than for ordinary subdiffusion, the jumps frequency is higher for the first process. The sequence of waiting times for a jump in both processes can be represented as follows
\begin{eqnarray}\label{eqVIII15}
\underbrace{(t_1,t_2-t_1,\ldots,t_n-t_{n-1})}_{ordinary\; subdif.}\Leftrightarrow\\ \underbrace{(g^{-1}(t_1),g^{-1}(t_2-t_1),\ldots,g^{-1}(t_n-t_{n-1}))}_{g-subdif.},\nonumber
\end{eqnarray}
with $t_i-t_{i-1}>g^{-1}(t_i-t_{i-1})$, $i=1,2,\ldots,n$. 

The $g$--Laplace transform of mean jumps number for $g$--subdiffusion $\left\langle n_g(t)\right\rangle=\sum_{n=1}^\infty nQ_{g,n}(t)$ is $\mathcal{L}_g[\left\langle n_g(t)\right\rangle](s)=\mathcal{L}_g[\psi_g(t)](s)/(s[1-\mathcal{L}_g[\psi_g(t)](s))]$ and the mean jumps frequency is $f_g(t)=d\left\langle n_g(t)\right\rangle/dt$. After calculation we get
\begin{equation}\label{eqVIII16}
\mathcal{L}_g[\left\langle n_g(t)\right\rangle](s)=\frac{1}{\tau s^{1+\alpha}},
\end{equation}
then
\begin{equation}\label{eqVIII17}
\left\langle n_g(t)\right\rangle=\frac{g^\alpha(t)}{\tau\Gamma(1+\alpha)},
\end{equation}
and 
\begin{equation}\label{eqVIII18}
f_g(t)=\frac{g'(t)}{\tau\Gamma(\alpha)g^{1-\alpha}(t)}.
\end{equation}

Figs. \ref{fig5}--\ref{fig8} presents exemplary plots of the time evolution of the average number of particle jumps and their frequency. Fig. \ref{fig5} shows the average number of particle jumps for $g$--subdiffusion for the various parameters $\gamma$, and they are compared with the plot obtained for ordinary subdiffusion. The plots are made on a linear scale. However, due to large differences in the function values these plots are also presented on a logarithmic scale in Fig. \ref{fig6}. The time evolutions of the jump frequency for the cases presented in Fig. \ref{fig5} are shown in Fig. \ref{fig7}. The frequencies increase much faster for lower values of the parameter $\gamma$. Fig. \ref{fig8} shows the influence of the parameter $\nu$ on the time evolution of mean number of jumps. For higher values of $\nu$, the function $a(t)$ faster goes to zero, thus $g$--subdiffusion faster goes to superdiffusion. This explains the greater values of $\left\langle n(t)\right\rangle$ for the greater values of $\nu$. 

Summarizing, the superdiffusion effect for $g$--subdiffusion is created by significantly increasing the jump frequency of the particle while the mean length of a single jump is finite.

\section{Final remarks\label{SecIX}}

We show that the $g$--subdiffusion equation with fractional time derivative with respect to another function $g$ can be used to describe a smooth transition from ordinary subdiffusion to superdiffusion. In some initial time interval the process is ordinary subdiffusion with parameters $\alpha$ and $D_\alpha$ described by the equation with the fractional Caputo time derivative. In the long time limit the process is superdiffusion with parameters $\gamma$ and $D_\gamma$ described by equation with the fractional spatial derivative. The transition from ordinary subdiffusion to superdiffusion is controlled by the function $a(t)$. We assume that fractional superdiffusion is well described by the $g$--subdiffusion equation when scaling properties of the Green's function for $g$--subdiffusion $P_g$ has scaling properties (at least approximately) such as the Green's function for fractional subdiffusion $P_\gamma$. The $g$--Laplace transform method can be used to solve the $g$--subdiffusion equation. The computational technique is then similar to that used to solve the ordinary subdiffusion equation by means of the ordinary Laplace transform method. 

The interpretation of the $g$--subdiffusion process which goes to superdiffusion is based on a particle jumps frequency which increases as a power function of time. Although the average jump length of the molecule is finite, in the long time limit this effect appears to be the same as the fractional superdiffusion effect generated by anomalously long molecule jumps with a constant jumps frequency. If the process were described by the ordinary subdiffusion equation Eq. (\ref{eqIII1}) with $\alpha>1$, the frequency of the particle jumps would increase with time \cite{koszt2019}. However, in this case, the process does not have a clear stochastic interpretation, as within the CTRW model the average waiting time for a molecule to jump is equal to zero when $\alpha>1$, see the discussion in Ref. \cite{koszt2019}. 

The Green's functions of $g$--subdiffusion equation $P_g$ and of fractional superdiffusion equation $P_\gamma$ are qualitatively different. However, despite the difference, the functions are equivalent in the long time limit. Paradoxically, a subdiffusion model describes superdiffusion. This result is also interesting from a mathematical point of view, as it exemplifies that solution to equation with a time fractional derivative may be asymptotically equivalent to solution to equation with a spatial fractional derivative; moreover, the fractional derivatives have different orders. 

Ordinary CTRW has been used to derive normal diffusion equation \cite{montroll1965,weiss1994}, ordinary subdiffusion and fractional superdiffusion equations \cite{mk,ks,mks,barkai2000,compte}, ultraslow diffusion (slow subdiffusion) equation \cite{denisov}, ordinary subdiffusion with reactions equation \cite{kl2014}, Cattaneo hyperbolic reaction--ordinary subdiffusion equation \cite{koszt2014}, fractional hyperbolic type Jeffreys equation \cite{awad2021}, and more. The $g$--subdiffusion equation can be derived by means of the $g$--CTRW model \cite{kd2021b}, which is a modified version of CTRW model. In the $g$--CTRW model the $g$--convolution and the $g$--Laplace transform are used, which makes the derivation of the Green's function similar to the derivation of the Green's function within the ordinary CTRW model. Since the $g$--CTRW model is based on a random walk model, the $g$--CTRW can also be used to derive similar equations to those mentioned above for the $g$--subdiffusion process. 

For sufficiently long times, $g$--subdiffusion is very close to fractional superdiffusion. The convergence of $g$--subdiffusion to fractional superdiffusion is determined by the function $a(t)$ in Eq. (\ref{eqVI5}). The question arises whether superdiffusion can be described by the $g$--subdiffusion equation in the entire time domain. This problem requires additional considerations such as an interpretation of the order $\alpha$ of $g$--subdiffusion equation. In our model, this order is defined for the initial stage of the process which is assumed to be ordinary subdiffusion. We add that normal diffusion is not excluded from the model. However, by selecting a function $a(t)$ that quickly goes to zero, the superdiffusion effect can be achieved for relatively short times. In such a case, the various equations derived from the $g$--CTRW model can be treated as equations describing superdiffusion.

As mentioned, for a sufficiently long time the Green's function for $g$--subdiffusion describes fractional superdiffusion. Even then, the $g$--subdiffusion model has some properties as ordinary subdiffusion model. This remark indicates additional possibilities in modeling superdiffusion processes. An example is the modeling of superdiffusion in a system with a thin partially permeable membrane. The determination of boundary conditions at the membrane plays a key role. 
Since the fractional superdiffusion equation is non--local in space, ``local'' boundary conditions are not considered for this equation. An example is the problem of boundary condition at a fully absorbing wall briefly discussed in Sec. \ref{SecII}. The conclusion is that the method of images and consequently the boundary condition at the absorbing wall $P_\gamma(x_M,t|x_0)=0$, where $x_M$ is a membrane location, should not be used for fractional superdiffusion. Instead, ``non-local'' boundary conditions for the absorbing wall, such as $P_\gamma(x,t|x_0)=0$ for $x\geq x_M$ (when $x_0<x_M$), can be considered. Another example is a boundary condition at a fully impermeable wall. In the stochastic model a ``wrapped around the boundary'' condition and a potential to hold a particle on the wall have been used \cite{dybiec2017}. However, even if such boundary conditions are formulated for the superdiffusion equation, they appear to be ineffective because of the difficulty of solving the equation.  Even greater is a problem to determine ``non-local'' boundary conditions at a reactive membrane and at a partially permeable wall where the particle which has passed through the membrane can pass back through it with some probability. For $g$--subdiffusion it is possible to derive boundary conditions at the membrane by means of the method used already for normal diffusion and ordinary subdiffusion, see for example Refs. \cite{koszt2019,weiss1994,koszt2015,koszt2021,grebenkov2010,grebenkov2020,korabel2010,korabel2011}. The methods mentioned above can even be applicable when $g$--subdiffusion is interpreted as superdiffusion.
 
The kind of diffusion is usually defined by time evolution of $\sigma^2$. However, this function does not always define the kind of process, see Refs. \cite{meroz2011,dybiec1,cherstvy2021}. In our considerations the diffusion process is characterized not only by $\sigma^2$, but also by the function $F$ Eq. (\ref{eqII12}). 
Superdiffusion is often defined in a simplified way as a process for which $\sigma^2_\gamma(t)\sim t^{2/\gamma}$ with $\gamma\in(1,2)$. Checking whether a process is superdiffusion is typically based on an experimental determination of the parameter $\gamma$ only. However, as argued in Sec. \ref{SecV}, for fractional superdiffusion there is $\sigma_\gamma^2(t)=\infty$ for $t>0$. $G$--subdiffusion equation describes superdiffusion when $g(t)\approx Et^{2/(\alpha\gamma)}$ for long times. In this case $\sigma_g^2$ is finite and reads  
\begin{equation}\label{eqIX1}
\sigma_g^2(t)=\frac{2D_\alpha E^\alpha}{\Gamma(1+\alpha)}t^{2/\gamma}.
\end{equation}
The model presented in this paper may also be used for a model of transition from ordinary subdiffusion to superdiffusion in which $\sigma^2(t)\sim t^\beta$ with $\beta\in(2,3)$ \cite{kelly2019}; however, then the condition that matches $P_g$ and $P_\gamma$ in the long time limit should be reconsidered.

\section*{Appendix}

The H--Fox is usually denoted as $H^{m\;n}_{p\;q}\left(z\left| \begin{array}{cc}
       \left[a_p,A_p\right] \\
        \left[b_q,B_q\right]
      \end{array}\right. \right)$, where
$0\leq m\leq q$, $0\leq n\leq p$, $[a_p,A_p]=(a_1,A_1),\ldots,(a_p,A_p)$, $[b_q,B_q]=(b_1,B_1),\ldots,(b_q,B_q)$, $A_i,B_j>0$ for $i=1,2,\ldots,p$, $j=1,2,\ldots,q$. Let $\Delta=\sum_{j=1}^q B_j-\sum_{j=1}^p A_j$ and $\mu =\sum_{j=1}^n A_j-\sum_{j=n+1}^p A_j+\sum_{j=1}^m B_j-\sum_{j=m+1}^q B_j$. The series representations of the H--Fox function are \cite{prud}
\begin{eqnarray}\label{a1}
H^{m\;n}_{p\;q}\left(z\left| \begin{array}{cc}
       \left[a_p,A_p\right] \\
        \left[b_q,B_q\right]
      \end{array}\right. \right)=\sum_{i=1}^m\sum_{l=0}^\infty \frac{(-1)^l z^{(b_i+l)/B_i}}{l!B_i}\\	
\times\frac{\prod_{j=1, j\neq i}^m\Gamma(b_j-(b_i+l)B_j/B_i)}{\prod_{j=n+1}^p\Gamma(a_j-(b_i+l)A_j/B_i)}\nonumber\\
\times\frac{\prod_{j=1}^n\Gamma(1-a_j+(b_i+l)A_j/B_i)}{\prod_{j=m+1}^q\Gamma(1-b_j+(b_i+l)B_j/B_i)},\nonumber
\end{eqnarray}
when $\Delta \geq 0$ and $B_l(b_j+r)\neq B_j(b_l+c)$, $j\neq l$, $j,l=1,2,\ldots,m$; $r,c=0,1,2,\ldots$, and
\begin{eqnarray}\label{a2}
H^{m\;n}_{p\;q}\left(z\left| \begin{array}{cc}
       \left[a_p,A_p\right] \\
        \left[b_q,B_q\right]
      \end{array}\right. \right)=\sum_{i=1}^n\sum_{l=0}^\infty\frac{(-1)^l z^{-(1-a_i+l)/A_i}}{l!A_i}\\
 \times\frac{\prod_{j=1, j\neq i}^n\Gamma(1-a_j-(1-a_i+l)A_j/A_i)}{\prod_{j=n+1}^p\Gamma(a_j+(1-a_i+l)A_j/A_i)}\nonumber\\
\times\frac{\prod_{j=1}^m\Gamma(b_j+(1-a_i+l)B_j/A_i)}{\prod_{j=m+1}^q\Gamma(1-b_j-(1-a_i+l)B_j/A_i)},\nonumber
\end{eqnarray}
when $\Delta <0$ and $A_l(1-a_j+r)\neq A_j(1-a_l+c)$, $j\neq l$, $j,l=1,2,\ldots,n$; $r,c=0,1,2,\ldots$. It is assumed that the "empty" product is equal to one. Eq. (\ref{a1}) has been involved in the considerations in Ref. \cite{mk}.

If $\Delta\leq 0$ or $\Delta>0$ and $\mu>0$, then \cite{mathai}
\begin{equation}\label{a3}
H^{m\;n}_{p\;q}\left(z\left| \begin{array}{cc}
       \left[a_p,A_p\right] \\
        \left[b_q,B_q\right]
      \end{array}\right. \right)=\mathcal{O}(z^d),\;|z|\rightarrow\infty,
\end{equation}
where $d={\rm min}_{1\leq j\leq n}[(a_j-1)/A_j]$.

For simplicity, we denote $\mathcal{F}[f(x)](k)=\tilde{f}(k)$ and $\mathcal{L}[f(t)](s)=\hat{f}(s)$, and assume $x_0=0$ (here $P_\gamma(x,t)\equiv P_\gamma(x,t|x_0)$).
Due to the following Fourier transform of the Riesz fractional derivative
\begin{equation}\label{a4}
\mathcal{F}\left[\frac{d^\gamma}{dx^\gamma}f(x)\right](k)=-|k|^\gamma\tilde{f}(k),
\end{equation}
the Fourier and the ordinary Laplace transforms of Eq. (\ref{eqV1}) provide
\begin{equation}\label{a5}
\tilde{\hat{P}}(k,s)=\frac{1}{s+D_\gamma |k|^\gamma}.
\end{equation}
Using the formula
\begin{equation}\label{a6}
\mathcal{L}^{-1}\left[\frac{1}{s+a}\right](t)={\rm e}^{-at},
\end{equation}
we get
\begin{equation}\label{a7}
\tilde{P}_\gamma(k,t)={\rm e}^{-Dt|k|^\gamma}.
\end{equation}
Since $\tilde{P}_\gamma(k,t)\equiv\tilde{P}_\gamma(-k,t)$, we have
\begin{equation}\label{a8} 
\mathcal{F}^{-1}[\tilde{P}_\gamma(k,t)](x)\equiv (1/\pi)\int_0^\infty {\rm cos}(kx)\tilde{P}_\gamma(k,t)dk. 
\end{equation}
The relations (\ref{a7}) and (\ref{a8}) provide
\begin{equation}\label{a9}
P_\gamma(x,t)=\frac{1}{\pi}\int_0^\infty {\rm cos}(kx){\rm e}^{-Dt|k|^\gamma}dk.
\end{equation}
Using the relation $\int_0^\infty x^\nu{\rm e}^{-ax^c} {\rm cos}(xy)dx=g_c(y)$, where $g_c(y)=(1/c)\sum_{i=0}^\infty (-1)^i a^{-(1+\nu+2i)/c}\Gamma((1+\nu+2i)/c)y^{2i}/(2i)!$, $\nu>-1$, $c>1$, and $a>0$ (see Eq. (3.56) in Ref. \cite{ober}) we get
\begin{equation}\label{a10}
\int_0^\infty {\rm cos}(kx){\rm e}^{-Dtk^\gamma}dk=\frac{1}{\gamma}\sum_{i=0}\frac{(-1)^i \Gamma((1+2i)/\gamma)x^{2i}}{(2i)!(Dt)^{(1+2i)/\gamma}}.
\end{equation}
We obtain
\begin{equation}\label{a11}
P(x,t)=\frac{1}{\pi\gamma}\sum_{i=0}^\infty \frac{(-1)^i \Gamma((1+2i)/\gamma) |x|^{2i}}{(2i)!(D_\gamma t)^{(1+2i)/\gamma}}.
\end{equation}
There are $(2i)!=(2i)!!(2i-1)!!$, where $(2i)!!=2\cdot 4\cdot\ldots\cdot 2i=2^i i!$, and $(2i-1)!!=1\cdot 3\cdot\ldots\cdot (2i-1)$.
Based on the following properties of Gamma function, $\Gamma(1+u)=u\Gamma(u)$, $\Gamma(1/2)=\sqrt{\pi}$, and $\Gamma(1+i)=i!$, where $i$ is a natural number, $i\in \mathcal{N}$, we get (see also Ch. 6 in \cite{abram})
\begin{equation}\label{a12}
\Gamma(i+1/2)=\frac{(2i-1)!!\sqrt{\pi}}{2^i},
\end{equation}
\begin{equation}\label{a13}
\Gamma(-i+1/2)=\frac{(-1)^i 2^i\sqrt{\pi}}{(2i-1)!!}.
\end{equation}
From the above equations we obtain
\begin{equation}\label{a14}
\Gamma(i+1/2)\Gamma(-i+1/2)=(-1)^i \pi,
\end{equation}
and 
\begin{equation}\label{a15}
(2i)!=\frac{\Gamma(i+1/2)i!4^i}{\sqrt{\pi}}.
\end{equation}
Combining Eqs. (\ref{a11}) and (\ref{a15}) we get Eq. (\ref{eqV4}). Due to Eq. (\ref{a1}), the series representation of Eq. (\ref{eqV3}) is Eq. (\ref{eqV4}). Thus, the solution to the superdiffusion equation is given by Eq. (\ref{eqV3}).

In Ref. \cite{mk} there has been shown that the Green's function of fractional superdiffusive equation is Eq. (\ref{eqV3a}).
Eqs. (\ref{eqV3a}) and (\ref{a1}) provide 
\begin{eqnarray}\label{a16}
P_\gamma(x,t)=\frac{1}{\gamma (D_\gamma t)^{1/\gamma}}\sum_{i=0}^\infty \left(-\frac{|x|}{(D_\gamma t)^{1/\gamma}}\right)^i \\ 
\times\frac{\Gamma((1+i)/\gamma)}{i!\Gamma((1-i)/2)\Gamma((1+i)/2)}.\nonumber
\end{eqnarray}
Since $1/\Gamma(-j)=0$ when $j\in\mathcal{N}$, the terms in the series on the right-hand side of Eq. (\ref{a16}) are non-zero only if $(1-i)/2\neq -j$, i.e. when $i=2j$. Putting $i=2j$ into the series and using Eqs. (\ref{a14}) and (\ref{a15}) we get
\begin{eqnarray}\label{a17}
P_\gamma(x,t)=\frac{1}{\gamma(D_\gamma t)^{1/\gamma}}\sum_{j=0}^\infty \left(-\frac{|x|}{(D_\gamma t)^{1/\gamma}}\right)^{2j} \\ 
\times\frac{\Gamma((1+2j)/\gamma)}{(2j)!\Gamma(-i+1/2)\Gamma(i+1/2)}\nonumber\\
=\frac{1}{\gamma\sqrt{\pi}(D_\gamma t)^{1/\gamma}}\sum_{j=0}^\infty \left(-\frac{|x|^2}{4(D_\gamma t)^{2/\gamma}}\right)^j \nonumber\\ 
\times\frac{\Gamma(1\gamma+2j/\gamma)}{j!\Gamma(j+1/2)}.\nonumber
\end{eqnarray}
Thus, Eq. (\ref{a17}) takes the form of Eq. (\ref{eqV4}). The Green's function derived in Refs. \cite{mainardi2001,mainardi2005} are expressed by Eq. (\ref{eqV3b}). From this equation and Eq. (\ref{a1}) we get Eq. (\ref{a16}). In conclusion, the functions Eqs. (\ref{eqV3}), (\ref{eqV3a}), and (\ref{eqV3b}) are equivalent to each other.

\end{document}